**Disconnection formation via segregation-induced grain boundary phase transitions**

Zuoyong Zhang and Chuang Deng*

*Department of Mechanical Engineering, University of Manitoba, Winnipeg, Manitoba R3T 5V6, Canada*

* Corresponding author: Chuang.Deng@umanitoba.ca

## Graphical abstract

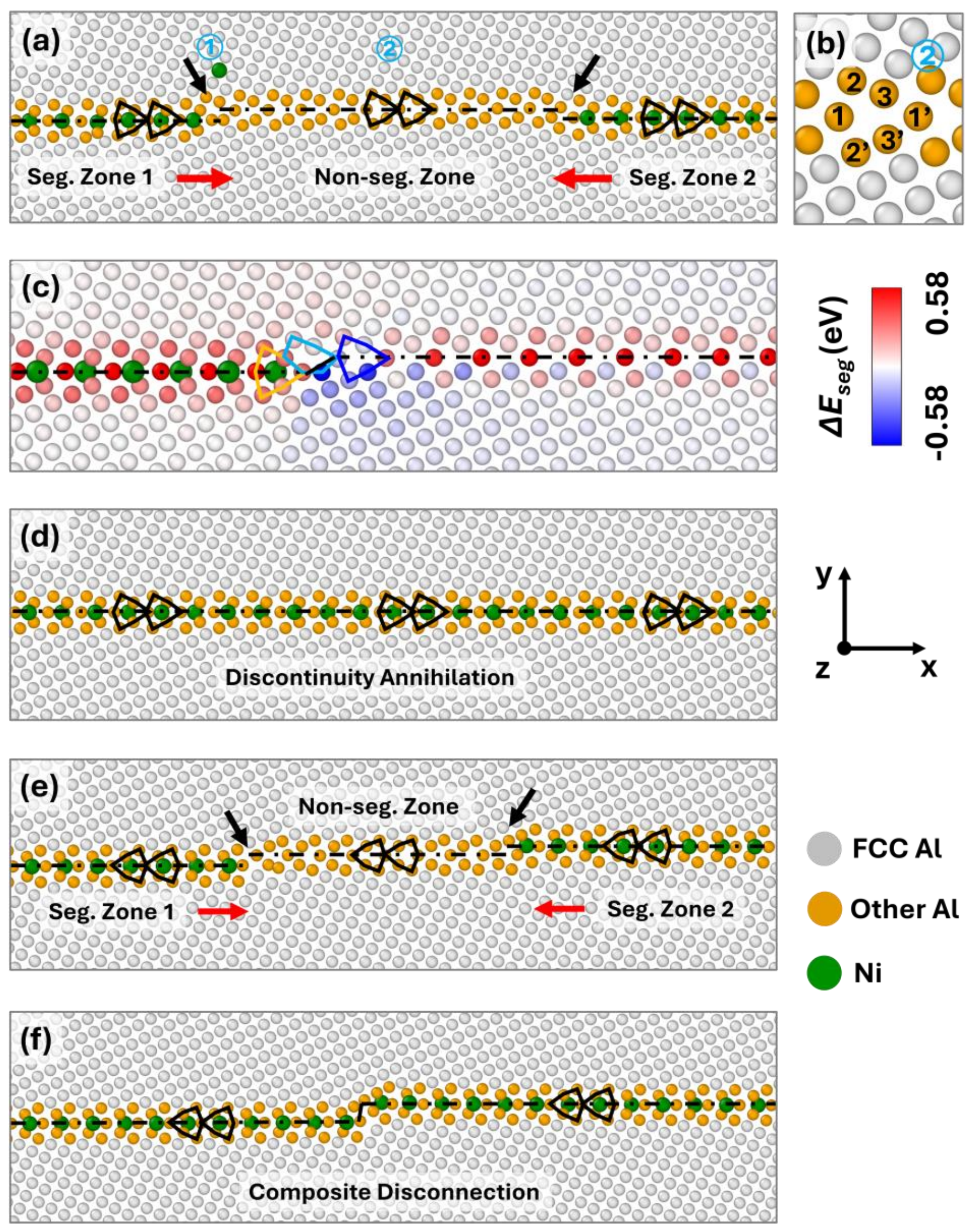

**Abstract**

Disconnections, long recognized as the key mediators of grain boundary (GB) kinetics in polycrystalline materials, have traditionally been understood to nucleate through thermal or mechanical activation. In this work, using atomistic simulations, we reveal a distinct nucleation mechanism driven exclusively by solute interstitial segregation across multiple substitutional binary alloy systems (e.g., Al-Ni, Al-Fe). This process exhibits zero-nucleation energy barriers, contrasting sharply with the nucleation mechanisms in pure systems. We identify states which are activated through segregation induced GB phase transitions: (i) isolated disconnections or phase junctions that promote GB migration and disappear with continuous segregation, and (ii) composite disconnections that are formed via two oppositely oriented isolated disconnections. The disconnections are mechanically robust, suppressing shear-coupled migration and instead resulting in GB amorphization and pure sliding under applied shear loading. The long-range stress fields associated with these composite disconnections further attract solute atoms and assist the nucleation of precipitates. These disconnections, absent in pure materials, follow unique nucleation pathways as confirmed through dichromatic pattern analysis and persist across different alloy chemistries and crystal structures. Our findings demonstrate that solute interstitial segregation provides a powerful and previously unrecognized pathway for barrier-free disconnection formation, thereby fundamentally extending current understanding of GB kinetics in alloy systems.

## 1 Introduction

Grain boundaries (GBs) and their kinetics play a crucial role in governing microstructure evolution and the physical properties of polycrystalline materials. In recent years, a unified theoretical framework centered on disconnections has emerged, identifying them as the fundamental defects mediating GB kinetics [1]. These GB disconnections are defined as line defects confined to the interfaces that possess both step and dislocation characters [2–4]. The step component is associated with the normal motion of the interface, whereas the dislocation component drives shear displacements between adjacent grains, thereby giving rise to shear-coupled GB migration in polycrystalline materials [1,5–8]. More recently, the dislocation character has also been linked to normal GB migration through generating driving forces such as internal stress, explaining the failure of the classical mean-curvature-flow model in describing grain growth [9].

According to the classical disconnection model [1], GB migration proceeds through two sequential processes: the nucleation of disconnections and their subsequent migration. The nucleation of a disconnection dipole requires overcoming a substantial activation energy barrier, typically accessible through thermal fluctuations or applied mechanical loading [5]. Once nucleated, the disconnection dipole can migrate by overcoming a much lower energy barrier [6]. In pristine materials, where nucleation occurs homogeneously, the overall kinetics of GB migration are generally limited by this nucleation process [10]. For a given GB structure, it is theoretically possible to enumerate all disconnection modes based on its dichromatic pattern [1]. This allows the corresponding nucleation barriers to be derived and, consequently, the migration kinetics to be predicted. The classical disconnection model offers a powerful framework for linking GB structure to migration behavior, a longstanding challenge due to the complexity of the five-dimensional GB parameter space, despite over a century of research [11,12].

Beyond mediating GB migration, disconnection motion also governs a range of other GB kinetic processes, including grain rotation [13–16], GB curvature evolution and roughening [1,17,18], and topological phase transitions at GBs [19]. Another type of GB discontinuity is the phase junction, which separates two distinct GB phases within the same boundary [20] and is associated with corresponding GB excess properties [21]. These phase junctions also possess the features of step heights and Burgers vectors, which is similar to regular GB disconnections in pure systems. While most studies on disconnection-mediated GB kinetics have focused on pure systems, impurity segregation introduces an additional mechanism for regulating GB kinetics, enabling phenomena such as segregation-induced GB phase transitions [22–24], diffusion-induced GB migration [25], and GB stabilization [26,27]. Recent studies have reported that pre-existing disconnections at twin boundaries can alter the solute segregation pattern due to the changes in local symmetry [28]. At low solute concentrations, disconnection dipoles tend to migrate, whereas higher concentrations can stabilize these dipoles through solute clustering at disconnection cores and provide pinning effects on the disconnection motion[29]. However, these investigations primarily examine segregation effects on twin boundary kinetics in systems where disconnections are artificially introduced. The role of solute segregation in the formation of GB disconnections, and its broader implications for GB kinetics, remains largely unexplored.

In this study, we report segregation-induced GB phase transitions that activate interesting disconnections which exhibit zero-nucleation energy barriers, as revealed by atomistic simulations. These disconnections exist in two distinct states: (i) isolated configurations at low solute concentrations and (ii) composite configurations when solute saturation is achieved at the GBs. Unlike the disconnections in pure systems, which are highly mobile and mediate shear-coupled GB motion, the segregation-activated disconnections are effectively immobile under applied shear loading and instead result in GB sliding. As the solute concentration increases further, stress fields associated with the stationary disconnection cores promote localized precipitation.

## 2 Methods

### *2.1 Atomistic simulations*

The atomistic simulations were conducted using the LAMMPS package [30]. OVITO was used for visualization [31], and the adaptive common neighbor analysis (a-CNA) method [32] was employed to analyze atomic structures. Hybrid molecular dynamics (MD)/Monte Carlo (MC) simulations were performed in the variance-constrained semi-grand-canonical (VC-SGC) ensemble [33] to sample the Al bicrystals in the Al-Ni system at 300 K and zero-pressure conditions. These bicrystals were constructed by replicating the original symmetric tilt grain boundary (STGB) structures from Ref. [34] between 5 and 51 times along the x and z directions, depending on the size of the initial GB unit cell. Periodic boundary conditions were applied in all directions of the simulation cell. Interatomic interactions were modeled using the embedded-atom method (EAM) potential for the Al-Ni system developed by Purja Pun and Mishin [35], which has been widely used in GB segregation studies [36–39]. In addition, EAM or other related potentials for the Al-Fe [40], Al-Cu [41], Ta-Cu [42], and W-Fe [43] systems were used for additional hybrid MD/MC simulations to further assess solute-induced GB behaviors.

A constant timestep of 2 fs was used throughout the hybrid MD/MC simulations. The bicrystal samples were initially relaxed using the isothermal-isobaric ensemble (NPT) under zero pressure at 300 K for 200 ps, which was sufficient to relax the pre-optimized samples. Then, in each MC cycle, the number of trial moves was set to 10% of the total atoms in the system at 300 K. Additionally, each MC cycle was separated by 10 MD steps at the same temperature to relax the samples after the atomic moves and chemical mixing, where the temperature was controlled using the canonical ensemble (NVT) and the pressure was precisely maintained to 0 bar by a Berendsen barostat [44]. In this work, we gradually increased the solute concentration in increments of 0.05 at.%. 10000 MC trials were attempted for each solute concentration. The Al-Ni system was selected as a representative substitutional alloy because hybrid MD/MC sampling at 300 K revealed pronounced interstitial segregation at the GBs. In a substitutional alloy, such interstitial segregation constitutes a form of GB phase transition in which solute atoms occupy the hollow interstitial sites within the kite structural units rather than substituting for host atoms on the kite lattice sites [45]. This interstitial segregation may also be accompanied by additional structural phase transitions and GB evolution in certain coincidence-site lattice (CSL) GBs [45]. These observations lead us to hypothesize that interstitial segregation in substitutional alloys may promote the nucleation of disconnections. For alloy systems without suitable EAM potentials, we employed the Metropolis algorithm [46] to perform the hybrid MD/MC simulations, during which solute concentrations were also gradually increased to the desired solute concentration with the interval of 0.05 at.%.

To evaluate the site occupancy preference, the solute segregation energy in Al is calculated using molecular statics (MS) simulations at 0 K under zero-pressure conditions as described in Ref. [37]. The iteration process employs a Ni atom to substitute a GB site, followed by an energy minimization using the conjugate gradient (CG) method. The segregation energy ($\Delta E_{seg}$) is defined by the energy difference between a solute occupying a GB site and a reference bulk site [47]:

$$\Delta E_{seg} = E_{GB}^{X} - E_{bulk}^{X}, \quad (1)$$

where the $E_{GB}^{X}$ is the system energy when a solute atom $X$ is occupying a GB site, and $E_{bulk}^{X}$ refers to the reference energy when a solute atom $X$ is located at the reference bulk site in the middle between two CSL GBs. It is worth noting that the single Ni atom introduced for the $\Delta E_{seg}$ calculations was constrained to remain at initial lattice position, while the surrounding solvent atoms were allowed to relax. This artificial constraint prevents the Ni atom from unintentionally migrating into interstitial sites during static energy minimization, analogous to the constrained-atom relaxation approaches commonly used in defect studies [48,49].

### *2.2 First-principles simulations*

Recent first-principles calculations [50–53] have revealed that some under-sized transition metal elements, such as Cu, Fe, and Ni, prefer to occupy the hollow sites within the Al STGBs. These studies inspire the interstitial site identification and improvement in GB segregation prediction [45]. However, the solute segregation path and how it would affect the GB evolution are still unclear. In this study, we employ first-principles simulations to investigate the solute segregation activated GB evolution and disconnection formation.

The density-functional theory (DFT) calculations were performed using the Vienna Ab-initio Simulation Package (VASP) [54,55], employing the projector augmented-wave (PAW) method [56]. The exchange-correlation potential was treated using the generalized gradient approximation (GGA) in the Perdew-Burke-Ernzerhof (PBE) form [57]. A cutoff energy of 350 eV and a Monkhorst-Pack k-point mesh of 7×5×3 was used for the Σ5(210) STGB. The convergence criteria were set to $10^{-6}$ eV for total energy and $10^{-2}$ eV/Å for atomic forces during the relaxation. The Σ5(210) STGB in Al was constructed following the method described in Ref. [58]. Before relaxing the STGB supercell, a bulk (4 ×4 ×4) supercell with 256 Al atoms was created and relaxed to equilibrium. The resulting lattice constant is 4.039 Å, which agrees well with the DFT calculation result of 4.041 Å [52] and experimental measurement of 4.046 Å [59]. Then, the initial STGB model, consisting of 160 Al atoms, was fully relaxed, as shown in Fig. 1. Subsequently, the 3 or 3′ sites were substituted with solute atoms. Then, the alloyed supercell was again fully relaxed to allow GB evolution and atomic rearrangement.

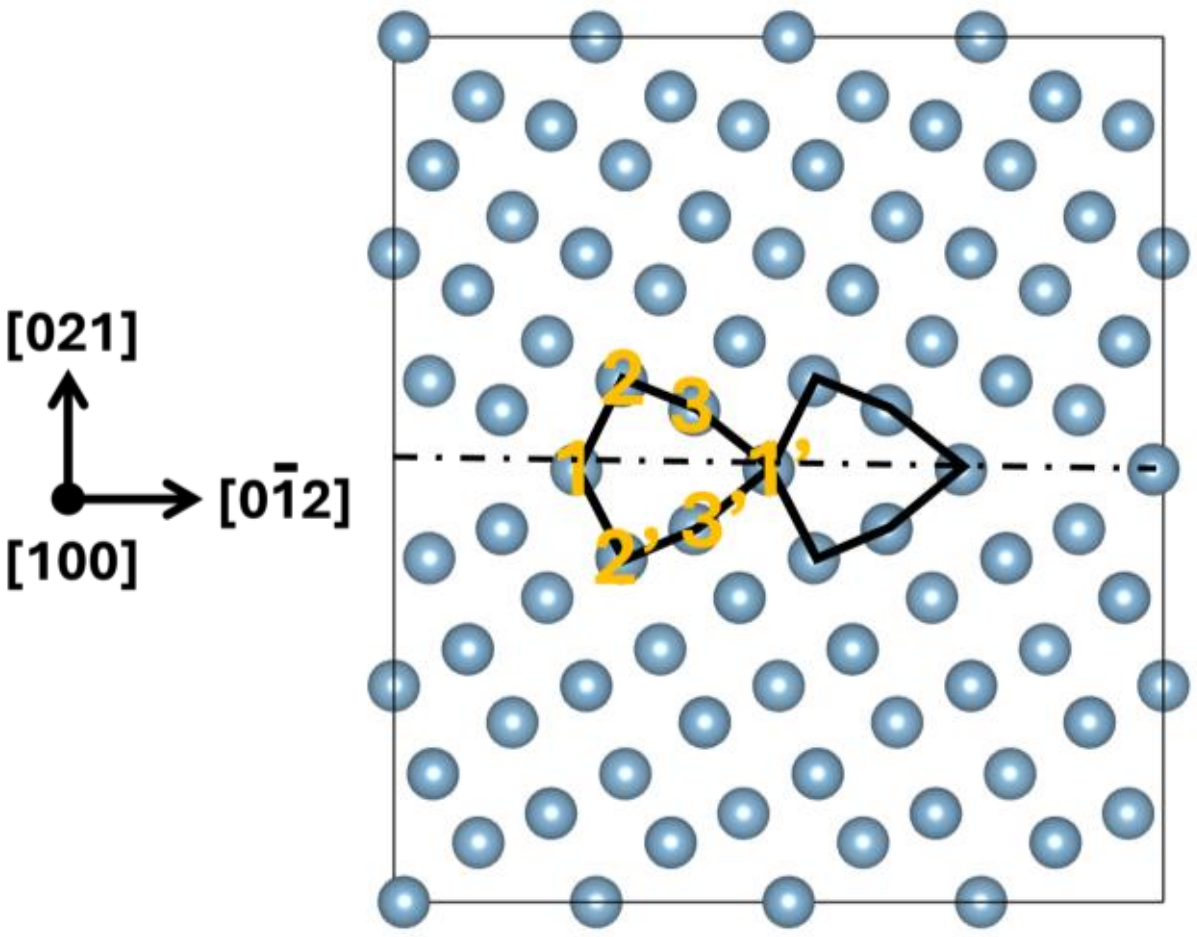

**Fig. 1** Relaxed atomic structure of the Σ5(210) STGB for the DFT simulations with 160 Al atoms and two Σ5(210) STGBs. The labeled 1, 2, 3, 1', 2', and 3' atomic sites are substitute GB sites within the GBs.

## 3 Results

### *3.1 Atomistic simulation results*

Figs. 2 and 3 illustrate the nucleation and evolution of two distinct disconnection states induced by interstitial segregation in a representative substitutional binary alloy system, Al-Ni. State (i) corresponds to isolated disconnections or phase junctions, which arise when solute atoms at the GBs are not yet saturated, and these disconnections are sensitive to solute concentrations and subsequently disappear as segregation continues. State (ii) corresponds to composite disconnections, which form once the GBs become solute-saturated. In this regime, two oppositely oriented isolated disconnections merge to produce a stable composite disconnection. We therefore use the terms "isolated" and "composite" disconnections to clearly distinguish these two regimes of GB structural evolution.

After hybrid MD/MC simulations, Ni atoms were found to segregate to those hollow interstitial sites within the kite structures of the Al Σ5(210) symmetric tilt GB (STGB), consistent with previous observations [45]. At low global solute concentrations (e.g., $X_{Ni}^{global}$ < 0.67 at.%), the STGB tends to separate into two regions: (i) segregation zone, where Ni atoms occupy interstitial sites, and (ii) non-segregation zone (region ②), which remains free of solute atoms, as shown in Fig. 2(a). Notably, a disconnection dipole emerges at the joints between these two regions, as highlighted by the black arrows in Fig. 2(a).

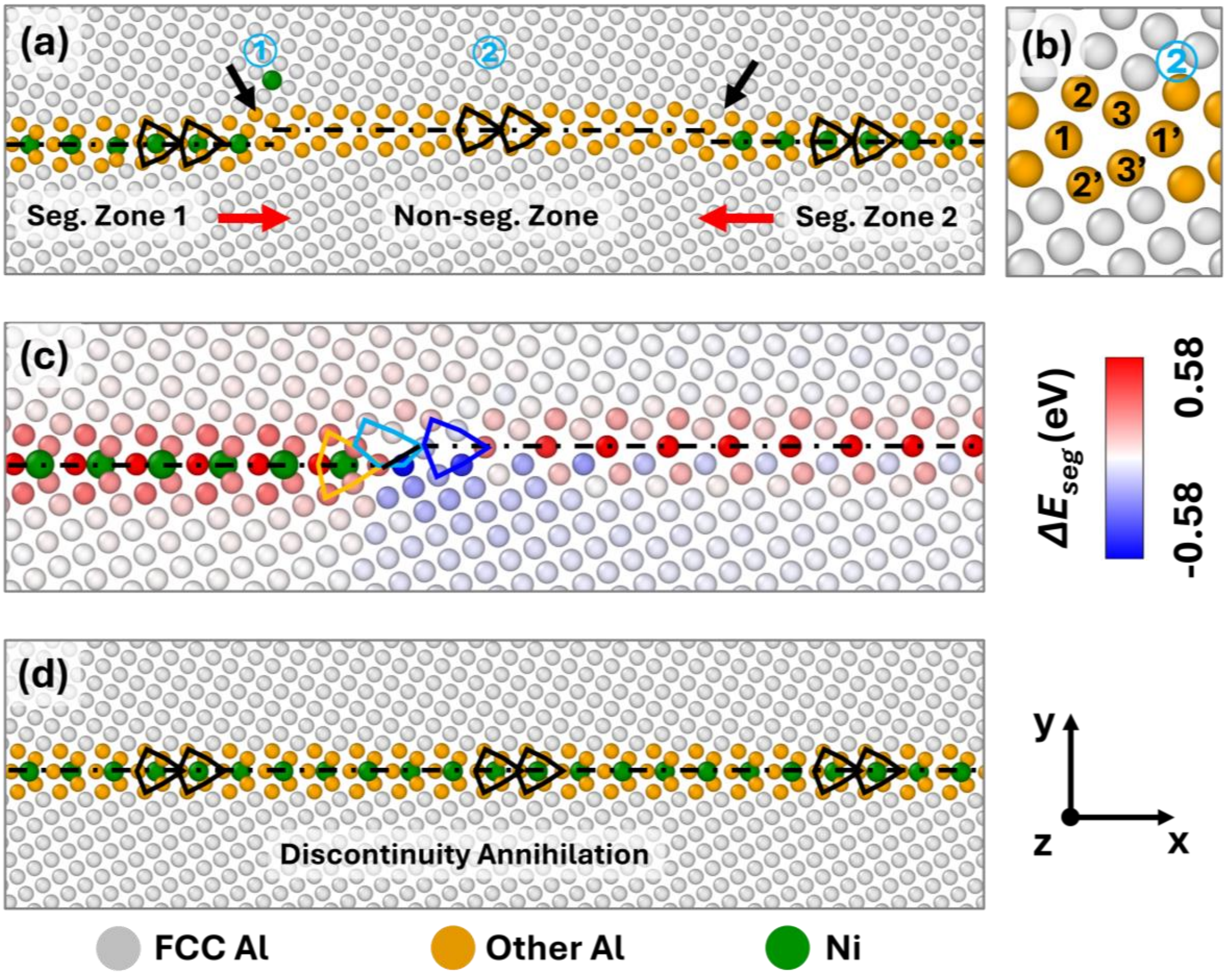

**Fig. 2** Isolated configurations observed in the Σ5(210) STGB after hybrid MD/MC simulations at 300 K in the Al-Ni system. (a) Nucleation of an isolated disconnection dipole at low solute concentration ($X_{Ni}^{global}$ < 0.67 at.%). (b) Atomic configuration at the non-segregated GB region ②. (c) Segregation energy $\Delta E_{seg}$ map of Ni in Al at a disconnection. (d) Disconnection annihilation at higher solute concentration ($X_{Ni}^{global}$ ≥ 0.67 at. %).

These isolated disconnections can be annihilated by increasing the solute concentration beyond 0.67 at.%, a threshold determined by the ratio of available interstitial sites within the two GBs of the Al bicrystal. The annihilation of such dipole results in an interstitially segregated flat GB, as shown in Fig. 2(d), driven by the continuous segregation of Ni atoms. Fig. 2(c) presents the segregation energy map of Ni atoms around the disconnection ① in Fig. 2(a). The segregated region exhibits a standard kite structure (orange), which connects to a highly distorted unsegregated kite (light blue), followed by a slightly distorted unsegregated kite (blue). The map indicates that Ni atoms preferentially segregate to site 3′ (as illustrated in Fig. 2(b)) within the light blue kite, which has the lowest segregation energy among all sites surrounding isolated disconnection ①. Upon segregation, the light blue kite transforms into a standard segregated kite, like the orange one to its left, through GB relaxation. This transformation subsequently distorts the adjacent blue kite, initiating a sequential process. A similar evolution occurs at the opposite disconnection of the dipole. Consequently, the overall segregation-induced annihilation follows the direction of the red arrows in Fig. 2(a), ultimately producing a flat GB in Fig. 2(d). Moreover, this interstitial segregation and discontinuity annihilation can induce GB migration downward. Due to the symmetry of the GB structure, upward GB migration may also occur if Ni atoms simultaneously segregate to site 3 at different regions of the GB.

Another scenario arises when solute atoms simultaneously segregate to sites 3 and 3' in the same GB. This leads to the formation of a disconnection "embryo", as illustrated in Fig. 3(a). Segregation to the upper sites (site 3) induces upward GB migration (Seg. Zone 2 in Fig. 3(a)), while segregation to the lower sites (site 3′) drives downward GB migration (Seg. Zone 1). The region between these two zones remains non-segregated, retaining the original kite structures characteristic of the pure system. The disconnections at the boundaries between these regions, marked by black arrows in Fig. 3(a), resemble the isolated configurations shown in Fig. 2(a). However, in contrast to the earlier case, further interstitial segregation drives the motion of these disconnections without leading to their annihilation. Instead, they merge to form a stable disconnection, as shown in Fig. 3(b). For clarity, only a single composite disconnection is shown here. Due to periodic boundary conditions, a corresponding opposite disconnection must exist elsewhere along the same GB. Additional Supplementary Movies 1 and 2 are provided to illustrate the formation of the two disconnection states during the hybrid MD/MC simulations. In these movies, red spheres represent solvent Al atoms, and blue spheres represent solute Ni atoms.

Additional hybrid molecular dynamics (MD)/kinetic Monte Carlo (kMC) simulations, performed using the method described in Ref. [60], further confirm that interstitial segregation of Ni atoms can trigger the formation of these composite disconnections, as shown in Supplementary Materials (SM) Fig. S1. Although this hybrid MD/kMC method still relies on the Metropolis acceptance criterion, it enables the simulation of atomic diffusion within the framework of MD simulations. Importantly, in this work, our focus is on capturing the thermodynamic aspects of the system rather than reproducing the exact kinetic behavior. Consequently, the method provides insights into equilibrium configurations and solute distributions, while the absolute timescales of diffusion events are not directly represented.

It is worth noting that for the state (i) discontinuities shown in Fig. 2(a), while analogous to phase junctions representing a partial phase transition between the $\alpha$ (unfilled-kite) and $\beta$ (filled-kite) phases, are primarily considered here in terms of their disconnection characteristics, specifically step heights and Burgers vectors. To highlight this distinction, we refer to them as isolated disconnections throughout the following discussion. However, those discontinuities presented in Fig. 3(b) are genuine disconnections, as the GB structures remain the same phase on both sides of these features. The interstitial segregation-induced GB phase transitions underlying these phenomena have been comprehensively reported and analyzed in our previous study [45]. Therefore, in the present work, we focus on the disconnection behavior itself to elucidate the mechanisms governing step formation and Burgers vector characteristics.

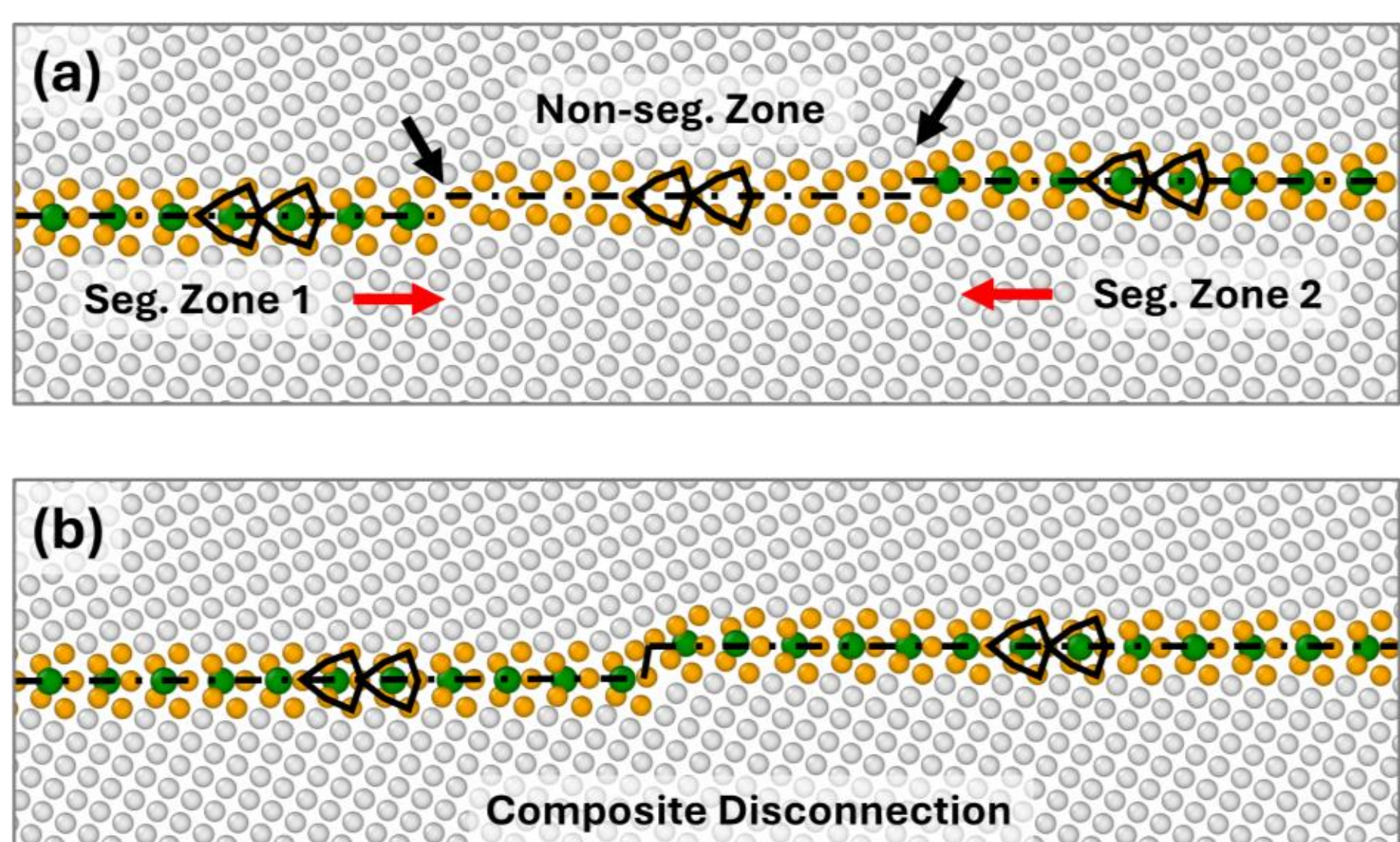

**Fig. 3** (a) Nucleation of state (ii) composite disconnections at low global Ni concentration. (b) Continuous interstitial segregation of Ni atoms finally forms a composite disconnection. Grey, orange, and green represent FCC Al, non-FCC Al, and Ni atoms, respectively.

### *3.2 DFT calculations*

Based on the MS simulation results shown in Fig. 2(a), Ni atoms were substituted into the 3 or 3′ sites within the kite structures, as illustrated in Fig. 4(a), (c), and (e). When the lower 3′ sites of two kite structures were substituted with Ni (Fig. 4(a)), the two kites migrated downwards and effectively “embraced” the Ni atoms into their interstitial sites, as shown in Fig. 4(b). An isolated disconnection dipole formed between the segregated and unsegregated regions. The adjacent unsegregated kites (outlined with dashed kites in Fig. 4(b)) were distorted by this disconnection, consistent with the hybrid MD/MC simulation results shown in Fig. 2(a) and (c). Because of the symmetry of the Σ5(210) STGB, its migration direction depends solely on the positions of the Ni atoms. This means that the substitution of two adjacent 3 sites in Fig. 1 leads to the upward migration of the half GB accordingly. These results confirm the observation of the nucleation of isolated disconnections from the hybrid MD/MC simulation results in Fig. 2(a) and the site-wise segregation energy map in Fig. 2(c).

When all 3′ sites were substituted with Ni atoms, as shown in Fig. 4(c), the entire STGB migrated downward by one atomic layer, and all Ni atoms remained at the interstitial sites within the kite structures after full relaxation, as shown in Fig. 4(d). This collective movement indicates a strong interaction between the solute atoms and the GB, whereby the presence of Ni atoms energetically favors a lower GB position. As a result, the solute atoms effectively "drag" the GB toward themselves, promoting the formation of a flat and energetically stabilized GB structure.

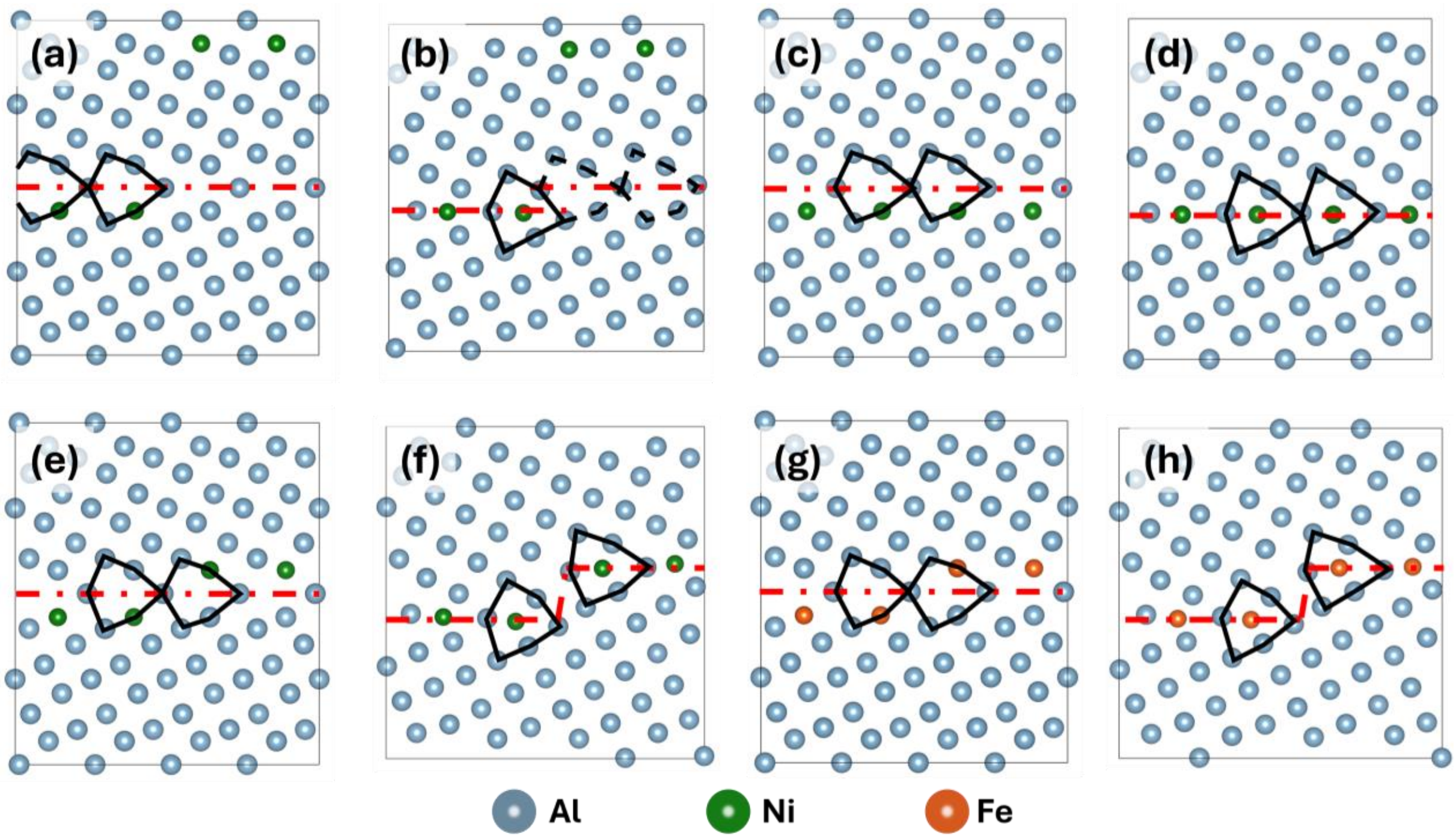

**Fig. 4** DFT simulation results for the Σ5(210) STGB in the Al-Ni and Al-Fe systems. (a), (c), (e), and (f) panels show the initial GB structure and solute positions before relaxation. (b), (d), (f), and (h) are the final GB structures after full relaxation.

However, with a sufficient number of solute atoms, if some Ni atoms occupy the 3′ sites while others simultaneously occupy the 3 sites, as shown in Fig. 4(e), this mixed occupation pattern disrupts the symmetry of the GB and ultimately leads to the nucleation of stable GB disconnections after full relaxation, as shown in Fig. 4(f). The resulting disconnections remain stable upon perturbation, providing direct atomic-scale evidence that solute segregation can promote and stabilize such disconnections. This finding further supports the formation mechanisms of the composite disconnections illustrated in Fig. 3. In addition, the Al-Fe system exhibits the same disconnections after relaxation, as shown in Figs. 4(g) and (h). The nucleation pathway and resulting defect structures closely mirror those observed in the Al-Ni system, indicating that the underlying segregation-driven mechanism is not chemistry-specific. This consistency across different solute species further supports the generality and robustness of the formation mechanism for composite disconnections.

It is worth noting that these DFT results indicate that the nucleation of both isolated and composite disconnections induced by solute interstitial segregation does not require any thermal activation. Instead, the process is governed entirely by the intrinsic interactions between the solute atoms and the GB. In other words, the segregation of solutes to specific GB sites is sufficient to spontaneously perturb the local atomic structure, lowering the energy barrier for disconnection nucleation to essentially zero. This highlights the critical role of solute-GB interactions in dictating GB structural transitions, even in the absence of external stimuli such as mechanical loading [5,61,62].

## 4 Discussion

### *4.1 GB behaviors under shear loading*

To compare the differences in GB kinetics between pristine and interstitially segregated GBs, shear deformation simulations were performed under applied shear loading at 0 K following the method described in Ref. [61]. The detailed simulation setups are provided in SM. Here, we would like to clarify that the segregated samples (both saturated and saturated/disconnected) were prepared artificially rather than through hybrid MD/MC simulations. Specifically, we first substituted the 3 or 3′ sites of the kite structures with solute atoms, followed by MS relaxation to obtain the flat saturated and disconnected GB configurations, respectively. This approach allows us to eliminate the influence of solute distribution and concentration, thereby isolating the effects of interstitial segregation and the presence of disconnections.

The results, summarized in Fig. 5, are based on bicrystal samples containing pristine, saturated, and saturated/disconnected GBs. In the pristine sample, once the shear stress reaches the critical value ($\tau_p^c =$ 2.02 GPa), the GB undergoes a shear-coupled migration event, accompanied by a sudden drop in shear stress, as illustrated in Fig. 5(a). With further shear displacement, the system enters a subsequent elastic loading regime until the shear stress again reaches the critical threshold, triggering repeated shear-coupled migration events, as shown in Fig. 5(b). This phenomenon is consistent with previously reported shear-coupled GB migration [5,61,62], but it exhibits a different shear-coupling factor, estimated as $\beta_0 = \Delta d_0/\Delta h_0 = 4.01$, where $\Delta d_0$ denotes the shear displacement along the GB plane and $\Delta h_0$ represents the associated GB migration distance, as illustrated in SM Fig. S3.

In the saturated sample, where all interstitial sites are fully occupied by Ni atoms, the critical shear stress for GB motion is significantly higher ($\tau_s^c = 3.66$ GPa) than that of the pristine case, as shown in Fig. 5(a). At this critical stress, the GB exhibits pure sliding without any accompanying normal motion, as shown in Fig. 5(c). This sliding behavior can be interpreted as a special disconnection mode, characterized by a finite Burgers vector but zero step height.

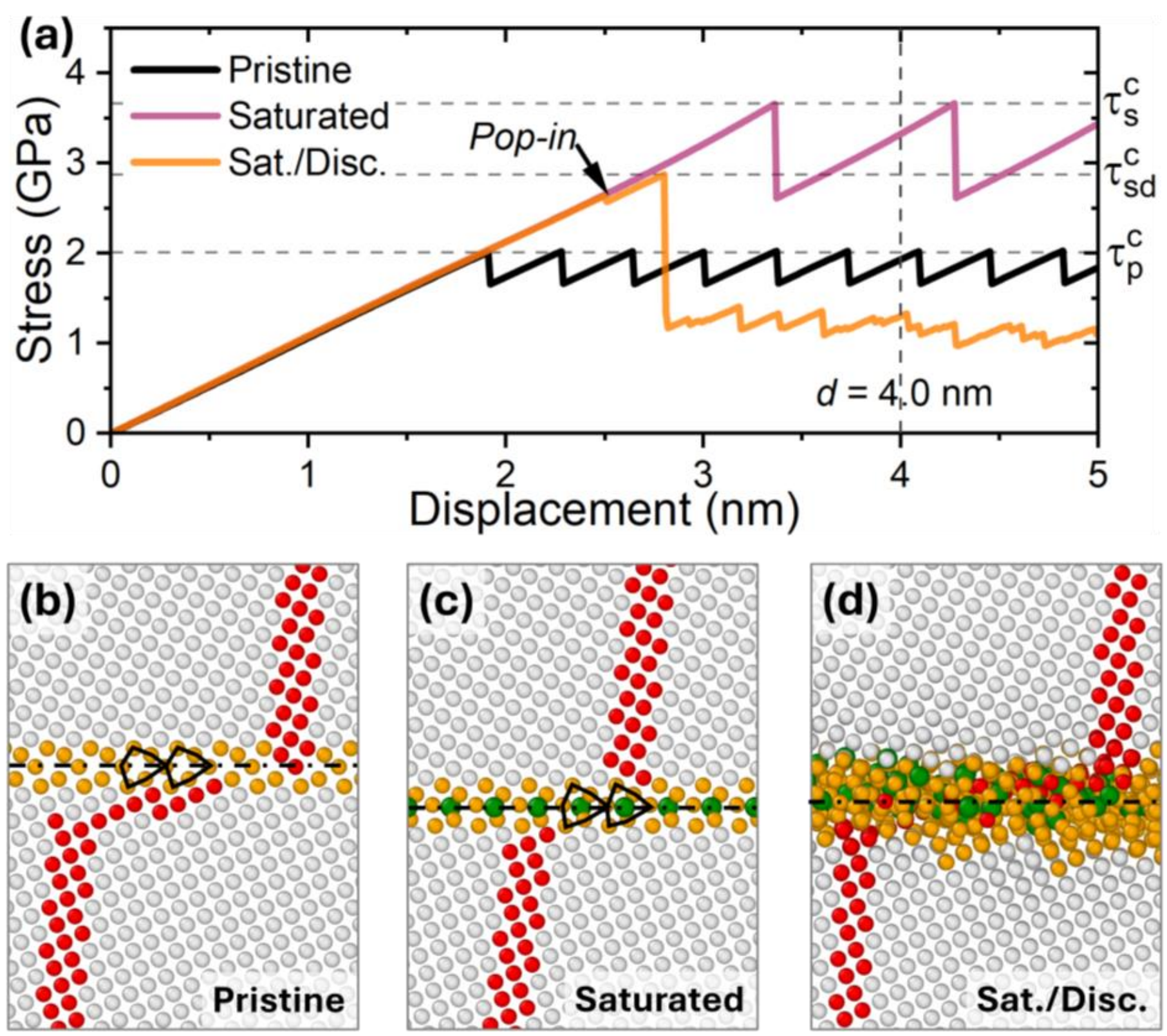

**Fig. 5** (a) Shear stress-displacement curves for bicrystals with pristine, saturated, and saturated/disconnected Σ5(210) STGBs. (b)-(d) GB behaviors at the displacement d = 4.0 nm. Red spheres serve as markers to track GB behavior under applied shear, while green spheres represent Ni atoms (enlarged for clarity).

In the saturated/disconnected case, where all interstitial sites are occupied by Ni atoms and a composite disconnection dipole is present, the critical shear stress ($\tau_{sd}^c = 2.86$ GPa) is higher than that of the pristine case but lower than that of the saturated case, as shown in Fig. 5(a). Notably, a pop-in event occurs at a lower shear stress ($\tau_{sd}^{pop-in} = 2.63$ GPa). This pop-in corresponds to a partial GB migration of the lower region of the interstitially segregated GB, as shown in SM Fig. S6. During this process, solute atoms transition from interstitial to substitutional sites of the kite structures, while the vacant interstitial sites become occupied by Al atoms. Following the pop-in event, the shear stress increases linearly with displacement until it reaches the critical point, at which the GB undergoes a sudden amorphization, like the GB structure shown in Fig. 5(d). This is followed by a transition to pure sliding, characterized by a significantly lower stress level, resulting in a flow-like shear stress-displacement response, as shown by the orange curve in Fig. 5(a). Nevertheless, the flow stress is significantly lower than the critical stress of pristine sample. The amorphization of the disconnected GB should be responsible for such softening behavior.

The immobile feature of saturated GBs, regardless of whether they contain composite disconnections, most likely governed by strong interactions between the interface and segregated solute atoms, namely solute drag or pinning effects on GB kinetics [27]. Even at finite temperatures, such as 300 K or 500 K, these heavily segregated GBs remain effectively immobile under applied shear loading (see SM Fig. S7). This stagnation demonstrates that such solute-GB interactions are sufficiently strong to suppress shear-coupled GB migration normal to the interface. Instead, the applied shear loading is accommodated primarily through pure interfacial sliding along the GB plane. Moreover, unlike disconnections in pure systems, which act as mediators of GB migration [7], disconnections induced by interstitial segregation are effectively immobile, highlighting an inherent distinction between segregation-induced disconnections and those in pure systems. Such behavior underscores the dominant role of solute drag in mediating GB kinetics in alloyed systems. However, a detailed examination of this effect lies beyond the primary scope of the present study and will be explored in future work.

These results demonstrate how interstitial segregation and preexisting composite disconnections profoundly affect the mechanical response and migration modes of GBs under shear loading. Specifically, interstitial Ni occupation increases the resistance to shear-driven GB motion and suppresses conventional shear-coupled migration, while promoting alternative deformation pathways such as pure sliding. Furthermore, the presence of a composite disconnection dipole introduces additional complexity by enabling solute-mediated structural rearrangements, such as stress-driven GB amorphization. These findings highlight the critical role of solute-GB interactions and disconnection structures in governing the stress thresholds, kinetic mechanisms, and overall deformation behavior of GBs under mechanical loading.

### *4.2 Dichromatic pattern analysis*

Dichromatic pattern analysis is a powerful tool to theoretically examining GB kinetics [1]. It provides a clear geometric framework for understanding how the relative displacement of adjacent grains gives rise to interfacial defects such as disconnections. This method has been successfully applied to elucidate the atomic-scale links between GB structure and mechanical response, thereby offering valuable insights into interfacial strength and deformation mechanisms [62].

In the present study, dichromatic pattern analysis is further employed to investigate how solute interstitial segregation affects the nucleation pathways of those disconnections. Fig. 6(a) compares two different disconnection nucleation pathways in defect-free Σ5(210) STGBs. The orange step height $h$ and Burgers vector **b** correspond the nucleation path of a conventional disconnection widely reported in the pristine Cu Σ5(210) STGB, characterized by a shear-coupling factor $\beta' = b/h = 2/3$ [5,62], where $\beta'$ denotes the shear-coupling factor and is introduced here to avoid confusion with the $\beta$ phase associated with GB complexions. The mechanism involves a three-layer GB migration normal to the boundary plane, during which the upper grain simultaneously shifts by two atomic layers relative to the lower grain along the GB plane. However, in the pristine Al Σ5(210) STGB, the disconnection nucleation pathway is highly sensitive to the choice of interatomic potentials. Shear simulations using some EAM potentials [40,63,64] and a recently developed machine-learning model [41] reproduce the same shear-coupling factor $\beta' = b/h = 2/3$, whereas other potentials [35,65,66] predict an alternative nucleation pathway. This alternative mechanism involves a single-layer normal migration and yields a higher shear-coupling factor $\beta_0' = b_0/h_0 = 4$, with the corresponding path highlighted by the dashed blue kite in Fig. 6(a).

The orange and blue (**b**, $h$) pairs in Fig. 6(b) and (c) illustrate the disconnection nucleation pathways induced by solute interstitial segregation, obtained using the EAM potential for the Al-Ni binary system [35]. Notably, both pathways involve single-layer GB migration but produce step heights of equal magnitude and opposite sign, such that $h_1 = -h_2 = d$, where $d$ is the distance between two adjacent atomic layers, as illustrated in Fig. 6(b). This behavior arises from the symmetry of the GB and the opposite initial positions of Ni interstitials: occupation of 3 sites yields a positive step height, while occupation of 3′ sites generates the opposite. The topological locations of the relevant interstitial sites are defined by the intersection points of the short-dashed black lines connecting sites 2-3′ and 2′-3 within the kite structures shown in Fig. 6(b) and (c). The solid arrows with open heads that link the current 3 or 3′ sites to the future interstitial sites represent the Burgers vectors, even though they do not directly connect two atoms. These arrows also illustrate the atomic displacement pathways associated with solute interstitial segregation. The dashed arrows with open heads indicate the step heights, as shown in Fig. 6(b) and (c). It is important to note that, in the dichromatic analysis, we neglect the contribution of the phase difference to the Burgers vectors. The actual components of the step heights and Burgers vectors for both the unstable and stable disconnections are provided in the Burgers circuit analysis shown in SM Fig. S8.

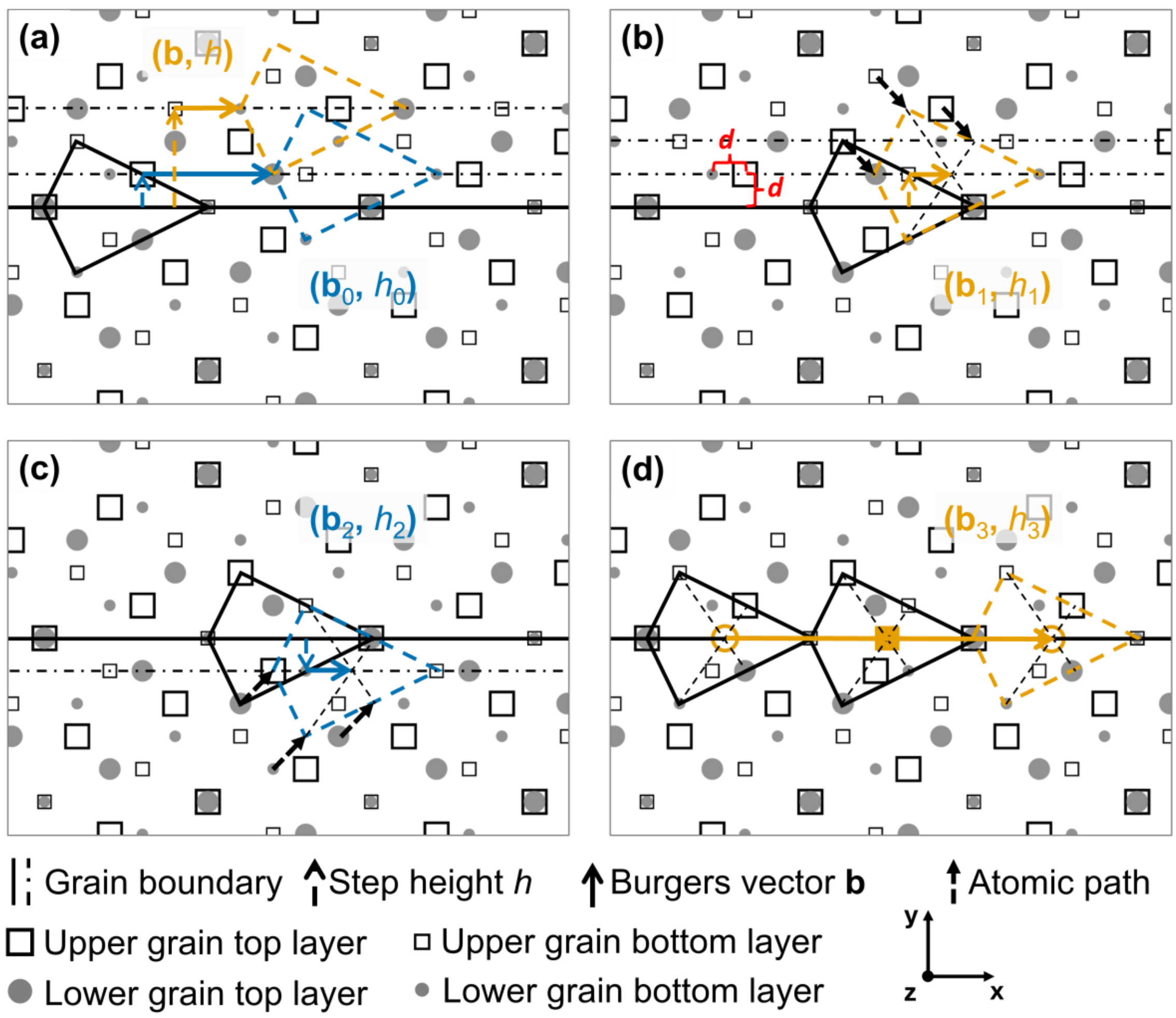

**Fig. 6** Dichromatic pattern analysis of an Al Σ5(210) STGB. (a) Comparison of the disconnection nucleation pathways for the pristine Σ5(210). (b) and (c) represent the Ni interstitial segregation-induced disconnection nucleation pathways. (d) Dichromatic pattern of the saturated Σ5(210) STGB (Fig. 2(c)) under applied shear loading. The black solid kites represent the original kite structures, while the dashed kites are new kites after nucleation.

Topological analysis reveals that both disconnections share the same Burgers vector: $\mathbf{b_1} = \mathbf{b_2} = 4\mathrm{d}/3$. As a result, their shear-coupling factors have equal magnitude but opposite signs, $\beta_1' = -\beta_2' = 4/3$, resemble the estimated values listed in SM Table S1. The disconnections formed via the pathways shown in Fig. 6(b) and (c) are individually unstable, as they arise solely from solute interstitial segregation. This means that if only one type of such disconnections is present, they will be eliminated as solute segregation continues, causing the GB to undergo a single-layer normal migration (either upper or lower). However, when both disconnections shown in Fig. 6(b) and (c) nucleate within the same GB, their interaction leads to the formation of stable disconnections. These findings are consistent with the hybrid MD/MC observations and DFT results, further reinforcing the mechanisms revealed by our multiscale analysis.

The corresponding dichromatic pattern analysis for the saturated sample under applied shear loading is illustrated by the orange arrow in Fig. 6(d), where the open circles and filled-square marks denote top and bottom interstitial sites, respectively. The analysis reveals a Burgers vector of $\mathbf{b_3} = 5\mathbf{b}$, which is intrinsically associated with the structural periodicity of the Σ5(210) STGB. This is further supported by the estimated elementary sliding distance, which spans approximately two GB kites, as shown in Fig. 5(c). Notably, the step height is zero, whereas the Burgers vector $\mathbf{b_3}$ is relatively large, indicating the pure sliding and a high resistance to shear-induced plastic deformation.

### *4.3 Precipitation at disconnections*

At disconnections, the presence of dislocation components inherently modifies the surrounding stress fields [1], and thereby influencing local segregation patterns [28,29]. This segregation behavior has been recognized as Cottrell atmosphere, where dislocation cores attract solute atoms to accumulate [67,68]. As shown in Fig. 7(a), for a given disconnection, the hydrostatic stresses exhibit opposite signs across GB and between its neighbors with opposite Burgers vectors. Here, negative and positive values reveal compressive and tensile regions, respectively. Consistent with the stress distribution, the segregation energy map in Fig. 7(b) follows the hydrostatic stress pattern. Compressive regions exhibit favorable segregation, reflected by negative segregation energies, while tensile sites correspond to solute anti-segregation, with positive segregation energies. This behavior aligns with the reported tendency of Ni atoms to preferentially segregate to compressive sites [37].

Once all interstitial sites are occupied by solute atoms, the GB can no longer accommodate additional solute atoms with positive segregation energies in either GB or near GB sites, as shown in Fig. 7(b). This behavior indicates the presence of repulsive solute-solute interactions, which stands in stark contrast to the solute-solute attraction commonly observed in nanocrystalline samples [39]. This discrepancy is likely attributed to the unique structural characteristics induced by interstitial solute segregation. As the solute concentration continues to increase, solute atoms are driven to disconnection cores, leading to local solute enrichment near these regions and promoting the formation of ordered solvent-solute phases (i.e., precipitates), as shown in Fig. 7(c). These precipitates exhibit a body-centered cubic (BCC) structure, as identified by common neighbor analysis shown in Fig. 7(d), indicating they are likely the B2 phase in the Al-Ni system. Composition analysis further confirms this, revealing an approximate Al-to-Ni atomic ratio of 1:1 within the BCC regions, consistent with that of the B2 phase [69]. Such precipitation behavior is characteristic of heterogeneous nucleation, which is a well-established mechanism for the formation of ordered intermetallic phases [70–72]. In this context, the formation of disconnections further promotes this process by providing energetically favorable sites for the heterogeneous nucleation of ordered Al–Ni phases.

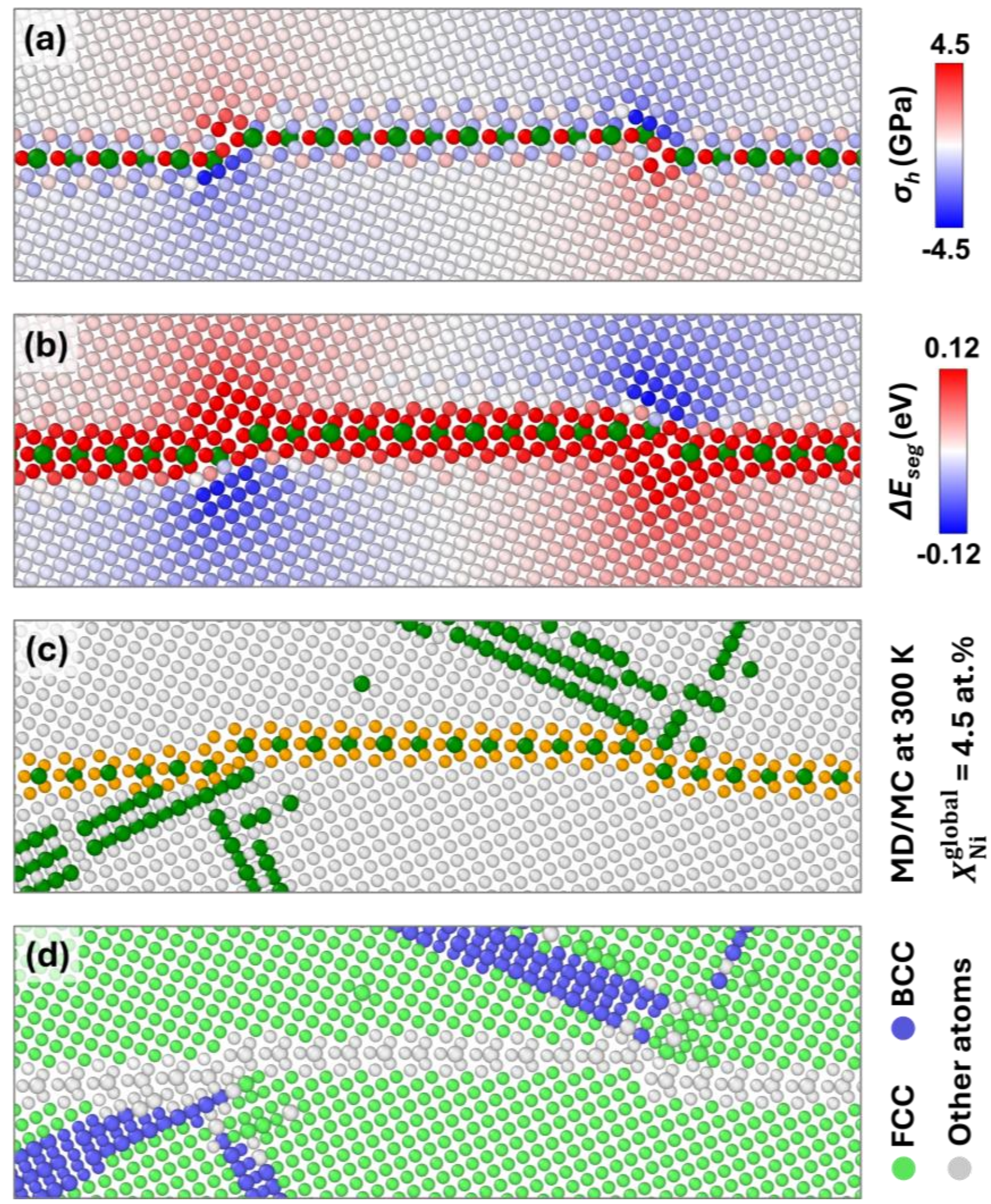

Fig. 7 (a) Hydrostatic stress ($\sigma_h$) map and (b) corresponding segregation energy ($\Delta E_{seg}$) map around the stable disconnection dipole at the Al-Ni Σ5(210) STGB. (c) Solute distribution at the disconnection dipole after hybrid MD/MC simulations at 300 K with a global solute concentration of 4.5 Ni at.%. (d) Common neighbor analysis of the atomic structure shown in (c). The green spheres in (a), (b), and (c) denote Ni atoms, which are enlarged for better visualization.

### *4.4 Disconnections in other systems*

Although the specific disconnection nucleation pathway in pristine Al Σ5(210) STGBs may vary depending on the interatomic potentials used, these differences do not influence the interstitial segregation or the subsequent formation of disconnections in the corresponding binary systems. For example, hybrid MD/MC simulations of the Al-Fe and Al-Cu systems reproduce similar composite disconnections triggered by Fe and Cu interstitial segregation in Al, as shown in Fig. 8(a) and (b), respectively. These disconnections are similar to the ones in the Al-Ni system, as shown in Fig. 3(b). This consistency persists despite the pristine Al Σ5(210) STGB exhibiting a shear-coupling factor of 2/3 when modeled using the Al-Fe EAM potential [40] and the UNEP machine-learning potential [41]. Therefore, these results demonstrate that the emergence of interstitial-segregation-induced disconnections is robust against the choice of interatomic potentials and reflects a general feature of solute interstitial-segregation-driven GB transformations rather than an artifact of a specific potential.

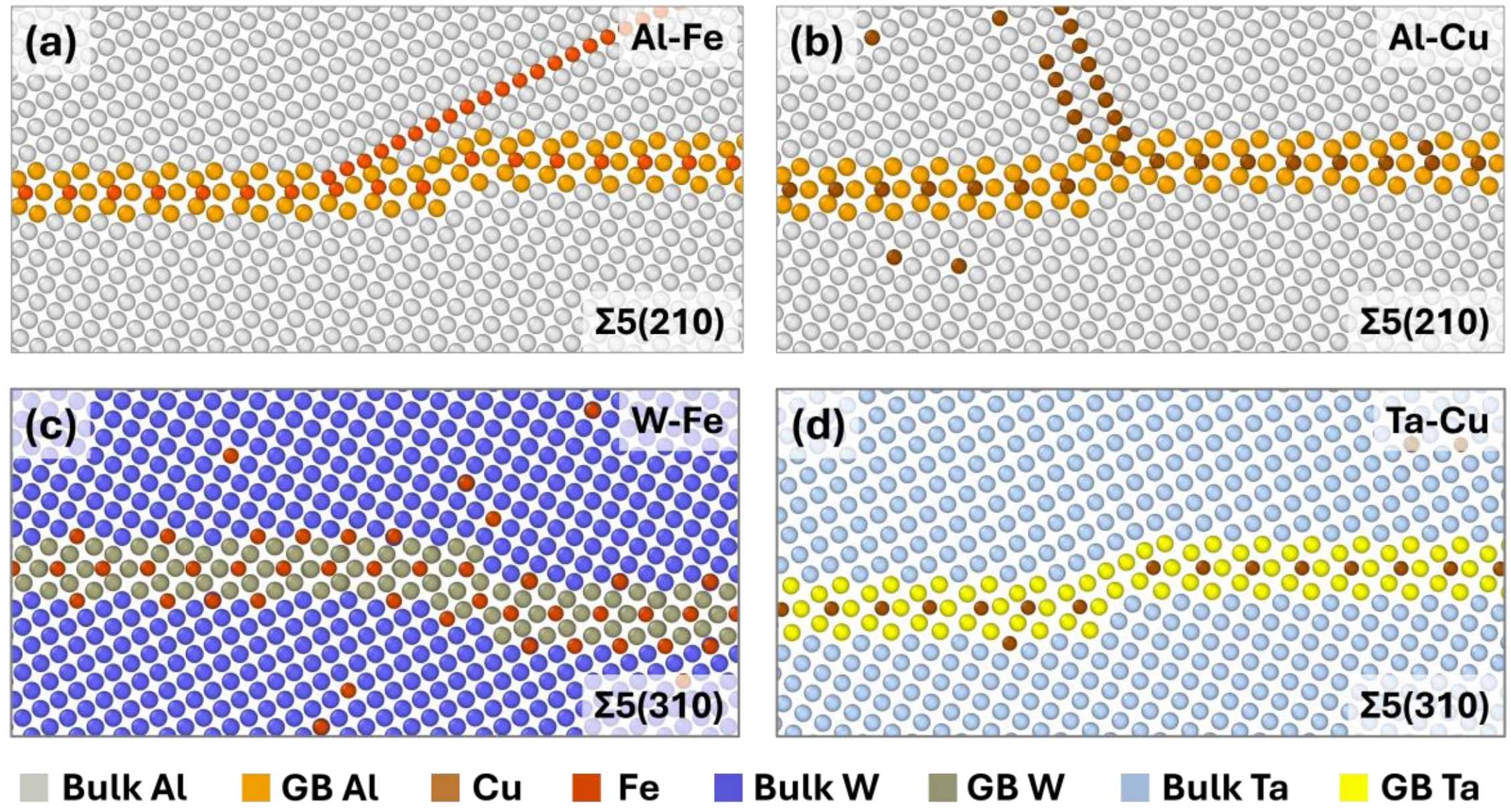

**Fig. 8** Permanent disconnections in the (a) Al-Fe, (b) Al-Cu, (c) W-Fe, and (d) Ta-Cu formed during the hybrid MD/MC simulations. Ordered solvent-solute structures (precipitates) can be observed in the Al-Fe and Al-Cu systems. Besides these face-centered cubic systems, GB phase transition-activated disconnections can also be observed in BCC systems including Ta-Cu and W-Fe within the Σ5(310) STGBs using their respective interatomic potentials [42,43], as shown in Fig. 8(c) and (d). Additionally, such disconnections were also identified within the Σ13(320) STGB in the Al-Ni system (see SM Fig. S9). In each of these binary systems, solute atoms preferentially occupy interstitial sites within the center of GB kite structures. In contrast, disconnections are absent in systems where solute atoms only occupy substitutional sites. These findings confirm that the nucleation of such disconnections is triggered by GB structural relaxation driven by solute interstitial segregation. Furthermore, the interstitial-segregation-activated disconnections exhibit a distinct relationship between the shear-coupling factor and GB structures compared to disconnections in pristine GBs [5], as analyzed in SM Fig. S10.

Although interstitial segregation is widely observed in substitutional binary alloys [45], it does not universally lead to disconnection formation. For example, low-angle ⟨001⟩ STGBs tend to promote precipitation, driven by stress fields from their dislocation array. We also did not observe GB disconnections in ⟨110⟩, ⟨111⟩ STGBs or asymmetric tilt GBs after hybrid MD/MC simulations under the same conditions. This may be attributed to the tendency of solute interstitials to form multi-layer segregation patterns or undergo structural transitions [45], which can balance the biased migration normal to the GB plane and thereby suppress disconnection nucleation. Currently, such interstitial-segregation-activated disconnections have only been observed in the high-angle ($\geq 53.13°$) ⟨001⟩ STGBs. A comprehensive exploration of solute-segregation-induced disconnection nucleation across other CSL and general GBs remains an open challenge, as these specific conditions may not commonly arise in polycrystalline materials [73].

## 4 Conclusion

In summary, we report the formation of GB disconnections induced by solute interstitial segregation in various substitutional binary alloys, including Al–Ni, Al–Fe, Al–Cu, Ta–Cu, and W–Fe, observed from hybrid MD/MC simulations at finite temperatures. The key conclusions are drawn as follows:

- Solute interstitial segregation activates two distinct types of disconnections: (i) isolated disconnections or phase junctions, which promote GB migration through GB evolution and relaxation, and eventually annihilate with continuous segregation; (ii) composite disconnections, which are formed through opposite GB migration driven by the isolated disconnections.
- First-principles and MS simulations confirm the interstitial-segregation-induced GB migration and formation of the two types of disconnections in both Al-Ni and Al-Fe systems. Importantly, the nucleation of such disconnections is barrierless, marking a critical distinction from the classical disconnections in pure systems. This zero-energy-barrier behavior is further supported by NEB calculations.
- The formed disconnections are mechanically robust and resist further normal GB migration but resulting in pure sliding along the GB plane under applied shear loading. The defect-free saturated sample reveals the highest critical stress to trigger sliding, whereas the disconnected/saturated sample undergoes GB amorphization before reaching critical stress, followed by sliding at a significantly reduced flow stress.
- The Dichromatic pattern analysis shows that solute segregation-induced disconnections can still be interpreted using classical disconnection model. Their step heights and Burgers vector components are independent of the employed interatomic potentials but are intrinsically determined by their GB structures.

- Upon saturation, the hydrostatic stress fields generated by composite disconnections promote additional solute accumulation, which may further drive the nucleation and growth of secondary phases, as observed in the Al–Ni, Al–Fe, and Al–Cu systems.

These findings advance our understanding of solute-governed GB kinetics and underscore the need for broader exploration of segregation-activated disconnections and their potential implications for alloy design.

## CRediT authorship contribution statement

**Zuoyong Zhang:** Writing – review & editing, Writing – original draft, Visualization, Validation, Software, Methodology, Investigation, Formal analysis, Data curation, Conceptualization. **Chuang Deng:** Writing – review & editing, Supervision, Software, Resources, Project administration, Methodology, Funding acquisition, Conceptualization.

## Declaration of Competing Interest

The authors declare that they have no known competing financial interests or personal relationships that could influence the work reported in this paper.

## Acknowledgement

This research was supported by the NSERC Discovery Grant (RGPIN-2019-05834), Canada, and the use of computing resources provided by the Digital Research Alliance of Canada. Z.Z. acknowledges financial support from the University of Manitoba Graduate Fellowship (UMGF). During the preparation of this manuscript, the authors used ChatGPT to improve its readability. The authors carefully reviewed and edited the manuscript following the use of this tool and took full responsibility for the content of the publication.

## Reference

[1] J. Han, S.L. Thomas, D.J. Srolovitz, Grain-boundary kinetics: A unified approach, Progress in Materials Science 98 (2018) 386–476. https://doi.org/10.1016/j.pmatsci.2018.05.004.

[2] J.P. Hirth, Dislocations, steps and disconnections at interfaces, Journal of Physics and Chemistry of Solids 55 (1994) 985–989. https://doi.org/10.1016/0022-3697(94)90118-X.

[3] J.P. Hirth, R.C. Pond, Steps, dislocations and disconnections as interface defects relating to structure and phase transformations, Acta Materialia 44 (1996) 4749–4763. https://doi.org/10.1016/S1359-6454(96)00132-2.

[4] J.P. Hirth, R.C. Pond, J. Lothe, Spacing defects and disconnections in grain boundaries, Acta Materialia 55 (2007) 5428–5437. https://doi.org/10.1016/j.actamat.2007.06.004.

[5] J.W. Cahn, Y. Mishin, A. Suzuki, Coupling grain boundary motion to shear deformation, Acta Materialia 54 (2006) 4953–4975. https://doi.org/10.1016/j.actamat.2006.08.004.

[6] H.A. Khater, A. Serra, R.C. Pond, J.P. Hirth, The disconnection mechanism of coupled migration and shear at grain boundaries, Acta Materialia 60 (2012) 2007–2020. https://doi.org/10.1016/j.actamat.2012.01.001.

[7] Q. Zhu, G. Cao, J. Wang, C. Deng, J. Li, Z. Zhang, S.X. Mao, In situ atomistic observation of disconnection-mediated grain boundary migration, Nat Commun 10 (2019) 156. https://doi.org/10.1038/s41467-018-08031-x.

[8] G.S. Rohrer, I. Chesser, A.R. Krause, S.K. Naghibzadeh, Z. Xu, K. Dayal, E.A. Holm, Grain Boundary Migration in Polycrystals, Annu. Rev. Mater. Res. 53 (2023) 347–369. https://doi.org/10.1146/annurev-matsci-080921-091511.

[9] C. Qiu, D.J. Srolovitz, G.S. Rohrer, J. Han, M. Salvalaglio, Why grain growth is not curvature flow, Proc. Natl. Acad. Sci. U.S.A. 122 (2025) e2500707122. https://doi.org/10.1073/pnas.2500707122.

[10] Y. Deng, C. Deng, Size and rate dependent grain boundary motion mediated by disconnection nucleation, Acta Materialia 131 (2017) 400–409. https://doi.org/10.1016/j.actamat.2017.04.018.

[11] G.L. Kelley, J. Winlock, On the restraint of exaggerated grain growth in critically strained metal, Journal of the Franklin Institute 201 (1926) 71–77. https://doi.org/10.1016/S0016-0032(26)91019-3.

[12] A. Bhattacharya, Y.-F. Shen, C.M. Hefferan, S.F. Li, J. Lind, R.M. Suter, C.E. Krill, G.S. Rohrer, Grain boundary velocity and curvature are not correlated in Ni polycrystals, Science 374 (2021) 189–193. https://doi.org/10.1126/science.abj3210.

[13] J.W. Cahn, J.E. Taylor, A unified approach to motion of grain boundaries, relative tangential translation along grain boundaries, and grain rotation, Acta Materialia 52 (2004) 4887–4898. https://doi.org/10.1016/j.actamat.2004.02.048.

[14] Y. Chen, H. Deng, Q. Zhu, H. Zhou, J. Wang, Direct observation of disconnection-mediated grain rotation, Scripta Materialia 252 (2024) 116279. https://doi.org/10.1016/j.scriptamat.2024.116279.

[15] C. Qiu, M. Salvalaglio, D.J. Srolovitz, J. Han, Disconnection flow–mediated grain rotation, Proc. Natl. Acad. Sci. U.S.A. 121 (2024) e2310302121. https://doi.org/10.1073/pnas.2310302121.

[16] Y. Tian, X. Gong, M. Xu, C. Qiu, Y. Han, Y. Bi, L.V. Estrada, E. Boltynjuk, H. Hahn, J. Han, D.J. Srolovitz, X. Pan, Grain rotation mechanisms in nanocrystalline materials: Multiscale observations in Pt thin films, Science 386 (2024) 49–54. https://doi.org/10.1126/science.adk6384.

[17] L. Zhang, J. Han, Y. Xiang, D.J. Srolovitz, Equation of Motion for a Grain Boundary, Phys. Rev. Lett. 119 (2017) 246101. https://doi.org/10.1103/PhysRevLett.119.246101.

[18] Y. Chen, J. Han, H. Deng, G. Cao, Z. Zhang, Q. Zhu, H. Zhou, D.J. Srolovitz, J. Wang, Revealing grain boundary kinetics in three-dimensional space, Acta Materialia 268 (2024) 119717. https://doi.org/10.1016/j.actamat.2024.119717.

[19] K. Chen, D.J. Srolovitz, J. Han, Grain-boundary topological phase transitions, Proc. Natl. Acad. Sci. U.S.A. 117 (2020) 33077–33083. https://doi.org/10.1073/pnas.2017390117.

[20] I.S. Winter, R.E. Rudd, T. Oppelstrup, T. Frolov, Nucleation of Grain Boundary Phases, Phys. Rev. Lett. 128 (2022) 035701. https://doi.org/10.1103/PhysRevLett.128.035701.

[21] T. Frolov, D.L. Medlin, M. Asta, Dislocation content of grain boundary phase junctions and its relation to grain boundary excess properties, Phys. Rev. B 103 (2021) 184108. https://doi.org/10.1103/PhysRevB.103.184108.

[22] T. Frolov, M. Asta, Y. Mishin, Segregation-induced phase transformations in grain boundaries, Phys. Rev. B 92 (2015) 020103. https://doi.org/10.1103/PhysRevB.92.020103.

[23] Z. Li, P. Zhao, W. Zhang, G. Xie, H. Duan, P. Zeng, S. Ma, J. Zhang, K. Du, Atomic unveiling of segregation-induced grain boundary structural transition in intermetallic compound, Acta Materialia (2025) 120830. https://doi.org/10.1016/j.actamat.2025.120830.

[24] V. Devulapalli, E. Chen, T. Brink, T. Frolov, C.H. Liebscher, Topological grain boundary segregation transitions, Science 386 (2024) 420–424. https://doi.org/10.1126/science.adq4147.

[25] R.W. Balluffi, J.W. Cahn, Mechanism for diffusion induced grain boundary migration, Acta Metallurgica 29 (1981) 493–500.

[26] T. Chookajorn, H.A. Murdoch, C.A. Schuh, Design of Stable Nanocrystalline Alloys, Science 337 (2012) 951–954. https://doi.org/10.1126/science.1224737.

[27] M. Hillert, Solute drag, solute trapping and diffusional dissipation of Gibbs energy, Acta Materialia 47 (1999) 4481–4505.

[28] C. Hu, S. Berbenni, D.L. Medlin, R. Dingreville, Discontinuous segregation patterning across disconnections, Acta Materialia 246 (2023) 118724. https://doi.org/10.1016/j.actamat.2023.118724.

[29] C. Hu, D.L. Medlin, R. Dingreville, Stability and Mobility of Disconnections in Solute Atmospheres: Insights from Interfacial Defect Diagrams, Phys. Rev. Lett. 134 (2025) 016202. https://doi.org/10.1103/PhysRevLett.134.016202.

[30] S. Plimpton, Fast Parallel Algorithms for Short-Range Molecular Dynamics, Journal of Computational Physics 117 (1995) 1–19. https://doi.org/10.1006/jcph.1995.1039.

[31] A. Stukowski, Visualization and analysis of atomistic simulation data with OVITO–the Open Visualization Tool, Modelling Simul. Mater. Sci. Eng. 18 (2009) 015012. https://doi.org/10.1088/0965-0393/18/1/015012.

[32] P.M. Larsen, S. Schmidt, J. Schiøtz, Robust structural identification via polyhedral template matching, Modelling and Simulation in Materials Science and Engineering 24 (2016) 055007.

[33] B. Sadigh, P. Erhart, A. Stukowski, A. Caro, E. Martinez, L. Zepeda-Ruiz, Scalable parallel Monte Carlo algorithm for atomistic simulations of precipitation in alloys, Phys. Rev. B 85 (2012) 184203. https://doi.org/10.1103/PhysRevB.85.184203.

[34] M.A. Tschopp, S.P. Coleman, D.L. McDowell, Symmetric and asymmetric tilt grain boundary structure and energy in Cu and Al (and transferability to other fcc metals), Integr Mater Manuf Innov 4 (2015) 176–189. https://doi.org/10.1186/s40192-015-0040-1.

[35] G.P. Purja Pun, Y. Mishin, Development of an interatomic potential for the Ni-Al system, Philosophical Magazine 89 (2009) 3245–3267. https://doi.org/10.1080/14786430903258184.

[36] M. Wagih, P.M. Larsen, C.A. Schuh, Learning grain boundary segregation energy spectra in polycrystals, Nat Commun 11 (2020) 6376. https://doi.org/10.1038/s41467-020-20083-6.

[37] Z. Zhang, C. Deng, Hydrostatic pressure-induced transition in grain boundary segregation tendency in nanocrystalline metals, Scripta Materialia 234 (2023) 115576. https://doi.org/10.1016/j.scriptamat.2023.115576.

[38] T.P. Matson, C.A. Schuh, A "bond-focused" local atomic environment representation for a high throughput solute interaction spectrum analysis, Acta Materialia 278 (2024) 120275. https://doi.org/10.1016/j.actamat.2024.120275.

[39] Z. Zhang, C. Deng, Grain boundary segregation prediction with a dual-solute model, Phys. Rev. Materials 8 (2024) 103605. https://doi.org/10.1103/PhysRevMaterials.8.103605.

[40] M.I. Mendelev, D.J. Srolovitz, G.J. Ackland, S. Han, Effect of Fe Segregation on the Migration of a Non-Symmetric Σ5 Tilt Grain Boundary in Al, J. Mater. Res. 20 (2005) 208–218. https://doi.org/10.1557/JMR.2005.0024.

[41] K. Song, R. Zhao, J. Liu, Y. Wang, E. Lindgren, Y. Wang, S. Chen, K. Xu, T. Liang, P. Ying, N. Xu, Z. Zhao, J. Shi, J. Wang, S. Lyu, Z. Zeng, S. Liang, H. Dong, L. Sun, Y. Chen, Z. Zhang, W. Guo, P. Qian, J. Sun, P. Erhart, T. Ala-Nissila, Y. Su, Z. Fan, General-purpose machine-learned potential for 16 elemental metals and their alloys, Nat Commun 15 (2024) 10208. https://doi.org/10.1038/s41467-024-54554-x.

[42] G.P. Purja Pun, K.A. Darling, L.J. Kecskes, Y. Mishin, Angular-dependent interatomic potential for the Cu–Ta system and its application to structural stability of nano-crystalline alloys, Acta Materialia 100 (2015) 377–391. https://doi.org/10.1016/j.actamat.2015.08.052.

[43] G. Bonny, N. Castin, J. Bullens, A. Bakaev, T.C.P. Klaver, D. Terentyev, On the mobility of vacancy clusters in reduced activation steels: an atomistic study in the Fe–Cr–W model alloy, J. Phys.: Condens. Matter 25 (2013) 315401. https://doi.org/10.1088/0953-8984/25/31/315401.

[44] H.J. Berendsen, J. van Postma, W.F. Van Gunsteren, A. DiNola, J.R. Haak, Molecular dynamics with coupling to an external bath, The Journal of Chemical Physics 81 (1984) 3684–3690.

[45] Z. Zhang, C. Deng, Grain boundary interstitial segregation in substitutional binary alloys, Acta Materialia 291 (2025) 121019. https://doi.org/10.1016/j.actamat.2025.121019.

[46] N. Metropolis, A.W. Rosenbluth, M.N. Rosenbluth, A.H. Teller, E. Teller, Equation of State Calculations by Fast Computing Machines, The Journal of Chemical Physics 21 (1953) 1087–1092. https://doi.org/10.1063/1.1699114.

[47] M. Wagih, C.A. Schuh, Spectrum of grain boundary segregation energies in a polycrystal, Acta Materialia 181 (2019) 228–237. https://doi.org/10.1016/j.actamat.2019.09.034.

[48] Y. Zheng, P. Yu, L. Zhang, Atomistic Study of the Interaction Nature of H-Dislocation and the Validity of Elasticity in Bcc Iron, Metals 13 (2023) 1267. https://doi.org/10.3390/met13071267.

[49] M.E. Foster, X.W. Zhou, Hydrogen isotope population near dislocations in zirconium from molecular dynamics, Heliyon 10 (2024) e32365. https://doi.org/10.1016/j.heliyon.2024.e32365.

[50] X.-Y. Liu, C.-L. Liu, L.J. Borucki, A new investigation of copper's role in enhancing Al–Cu interconnect electromigration resistance from an atomistic view, Acta Materialia 47 (1999) 3227–3231. https://doi.org/10.1016/S1359-6454(99)00186-X.

[51] D. Zhao, O.M. Løvvik, K. Marthinsen, Y. Li, Segregation of Mg, Cu and their effects on the strength of Al Σ5 (210)[001] symmetrical tilt grain boundary, Acta Materialia 145 (2018) 235–246. https://doi.org/10.1016/j.actamat.2017.12.023.

[52] X. Zhang, L. Zhang, Z. Zhang, X. Huang, Effect of solute atoms segregation on Al grain boundary energy and mechanical properties by first-principles study, Mechanics of Materials 185 (2023) 104775. https://doi.org/10.1016/j.mechmat.2023.104775.

[53] J. Tang, Y. Wang, Y. Jiang, J. Yao, H. Zhang, Solute Segregation to Grain Boundaries in Al: A First-Principles Evaluation, Acta Metall. Sin. (Engl. Lett.) 35 (2022) 1572–1582. https://doi.org/10.1007/s40195-022-01383-w.

[54] G. Kresse, J. Furthmüller, Efficiency of ab-initio total energy calculations for metals and semiconductors using a plane-wave basis set, Computational Materials Science 6 (1996) 15–50. https://doi.org/10.1016/0927-0256(96)00008-0.

[55] G. Kresse, J. Furthmüller, Efficient iterative schemes for *ab initio* total-energy calculations using a plane-wave basis set, Phys. Rev. B 54 (1996) 11169–11186. https://doi.org/10.1103/PhysRevB.54.11169.

[56] P.E. Blöchl, Projector augmented-wave method, Phys. Rev. B 50 (1994) 17953–17979. https://doi.org/10.1103/PhysRevB.50.17953.

[57] J.P. Perdew, K. Burke, M. Ernzerhof, Generalized Gradient Approximation Made Simple, Phys. Rev. Lett. 77 (1996) 3865–3868. https://doi.org/10.1103/PhysRevLett.77.3865.

[58] D.L. Olmsted, S.M. Foiles, E.A. Holm, Survey of computed grain boundary properties in face-centered cubic metals: I. Grain boundary energy, Acta Materialia 57 (2009) 3694–3703. https://doi.org/10.1016/j.actamat.2009.04.007.

[59] W.P. Davey, Precision Measurements of the Lattice Constants of Twelve Common Metals, Phys. Rev. 25 (1925) 753–761. https://doi.org/10.1103/PhysRev.25.753.

[60] J.P. Tavenner, M.I. Mendelev, J.W. Lawson, Molecular dynamics based kinetic Monte Carlo simulation for accelerated diffusion, Computational Materials Science 218 (2023) 111929. https://doi.org/10.1016/j.commatsci.2022.111929.

[61] A. Rajabzadeh, F. Mompiou, M. Legros, N. Combe, Elementary Mechanisms of Shear-Coupled Grain Boundary Migration, Phys. Rev. Lett. 110 (2013) 265507. https://doi.org/10.1103/PhysRevLett.110.265507.

[62] Y. Deng, C. Deng, Atomic link between the structure and strength of grain boundaries subject to shear coupling, Phys. Rev. Materials 3 (2019) 010601. https://doi.org/10.1103/PhysRevMaterials.3.010601.

[63] M.I. Mendelev, M.J. Kramer, C.A. Becker, M. Asta, Analysis of semi-empirical interatomic potentials appropriate for simulation of crystalline and liquid Al and Cu, Philosophical Magazine 88 (2008) 1723–1750. https://doi.org/10.1080/14786430802206482.

[64] V.V. Zhakhovskii, N.A. Inogamov, Yu.V. Petrov, S.I. Ashitkov, K. Nishihara, Molecular dynamics simulation of femtosecond ablation and spallation with different interatomic potentials, Applied Surface Science 255 (2009) 9592–9596. https://doi.org/10.1016/j.apsusc.2009.04.082.

[65] R.R. Zope, Y. Mishin, Interatomic potentials for atomistic simulations of the Ti-Al system, Phys. Rev. B 68 (2003) 024102. https://doi.org/10.1103/PhysRevB.68.024102.

[66] Y. Mishin, D. Farkas, M.J. Mehl, D.A. Papaconstantopoulos, Interatomic potentials for monoatomic metals from experimental data and *ab initio* calculations, Phys. Rev. B 59 (1999) 3393–3407. https://doi.org/10.1103/PhysRevB.59.3393.

[67] A.H. Cottrell, B.A. Bilby, Dislocation Theory of Yielding and Strain Ageing of Iron, Proc. Phys. Soc. A 62 (1949) 49–62. https://doi.org/10.1088/0370-1298/62/1/308.

[68] Y. Mishin, J.W. Cahn, Thermodynamics of Cottrell atmospheres tested by atomistic simulations, Acta Materialia 117 (2016) 197–206. https://doi.org/10.1016/j.actamat.2016.07.013.

[69] M. Settem, On the structural analysis of ordered B2 AlNi nanoparticles obtained using freezing simulations, Intermetallics 106 (2019) 115–123. https://doi.org/10.1016/j.intermet.2018.12.018.

[70] D. Turnbull, Kinetics of Heterogeneous Nucleation, The Journal of Chemical Physics 18 (1950) 198–203. https://doi.org/10.1063/1.1747588.

[71] A.L. Greer, Overview: Application of heterogeneous nucleation in grain-refining of metals, The Journal of Chemical Physics 145 (2016) 211704. https://doi.org/10.1063/1.4968846.

[72] W. Hao, S. Ma, Z. Dong, Y. Hu, L. Wang, H. Chen, Q. Xu, H. Dong, Atomic characteristics of heterogeneous nucleation during solidification of aluminum alloys: A critical review, Materials Genome Engineering Advances 3 (2025) e65. https://doi.org/10.1002/mgea.65.

[73] M. Wagih, C.A. Schuh, Viewpoint: Can symmetric tilt grain boundaries represent polycrystals?, Scripta Materialia 237 (2023) 115716. https://doi.org/10.1016/j.scriptamat.2023.115716.

**Supplementary Materials for**

# Disconnection formation via segregation-induced grain boundary phase transitions

Zuoyong Zhang and Chuang Deng*

*Department of Mechanical Engineering, University of Manitoba, Winnipeg, Manitoba R3T 5V6, Canada*

*Corresponding author: chuang.deng@umanitoba.ca

## *Kinetic Monte Carlo simulation details*

The simulations were carried out using the hybrid molecular dynamics (MD)/kinetic Monte Carlo (kMC) method described in Ref. [1], in which MC trials are limited to swapping atoms with their nearest neighbors. This approach enables accelerated sampling of solute diffusion. The simulation box contains 93,000 atoms, with 2 at.% of Al atoms randomly substituted by Ni atoms. The hybrid MD/kMC simulations were then performed at 300 K and zero pressure for a total duration of 20 *ns*, followed by cooling the system to 0 K.

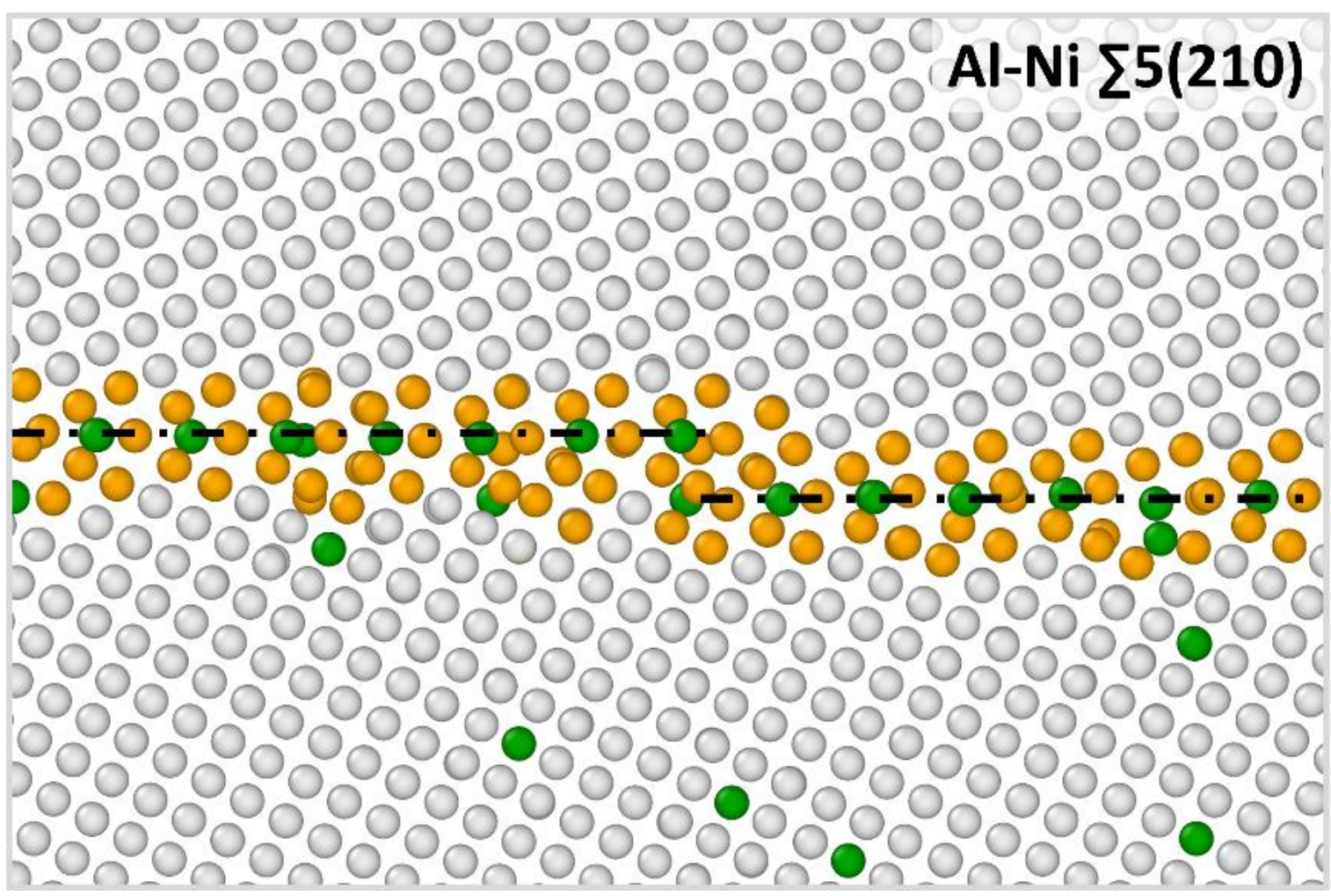

**Fig. S1** Disconnection formed after kMC simulations for the duration of 20 *ns* in the Al-Ni system.

### *Shear deformation simulation details*

To compare the grain boundary (GB) motions among the pristine symmetry tilt grain boundary (STGB), defect-free saturated GB, and saturated GB but with a disconnection dipole under shear stresses, shearing simulations were performed at 0 K following the method described in Ref. [2]. Periodic boundary conditions were applied along $x$- and $z$-directions, while leaving y-direction as free. Each bicrystal was first optimized using the conjugate gradient (CG) method. Then, the slabs were translated relatively to each other in the $x$-direction by small incremental displacements (0.05 Å) were applied to the top and bottom slabs in opposite directions (along $x$-axis), parallel to the GB plane. After each increment, the system energy was minimized using the CG method. The simulation cell layout and initial GB configurations of the samples are shown in Fig. S2. All samples have identical dimensions of $19.02 \times 27.05 \times 4.05\, nm^3$. It is worth noting that the saturated and saturated/disconnection samples generated through molecular statics (MS) simulations, during which the 3 or 3′ Al atoms were manually replaced by Ni atoms to avoid potential defects that might be introduced by hybrid molecular dynamics (MD)/Monte Carlo (MC) simulations.

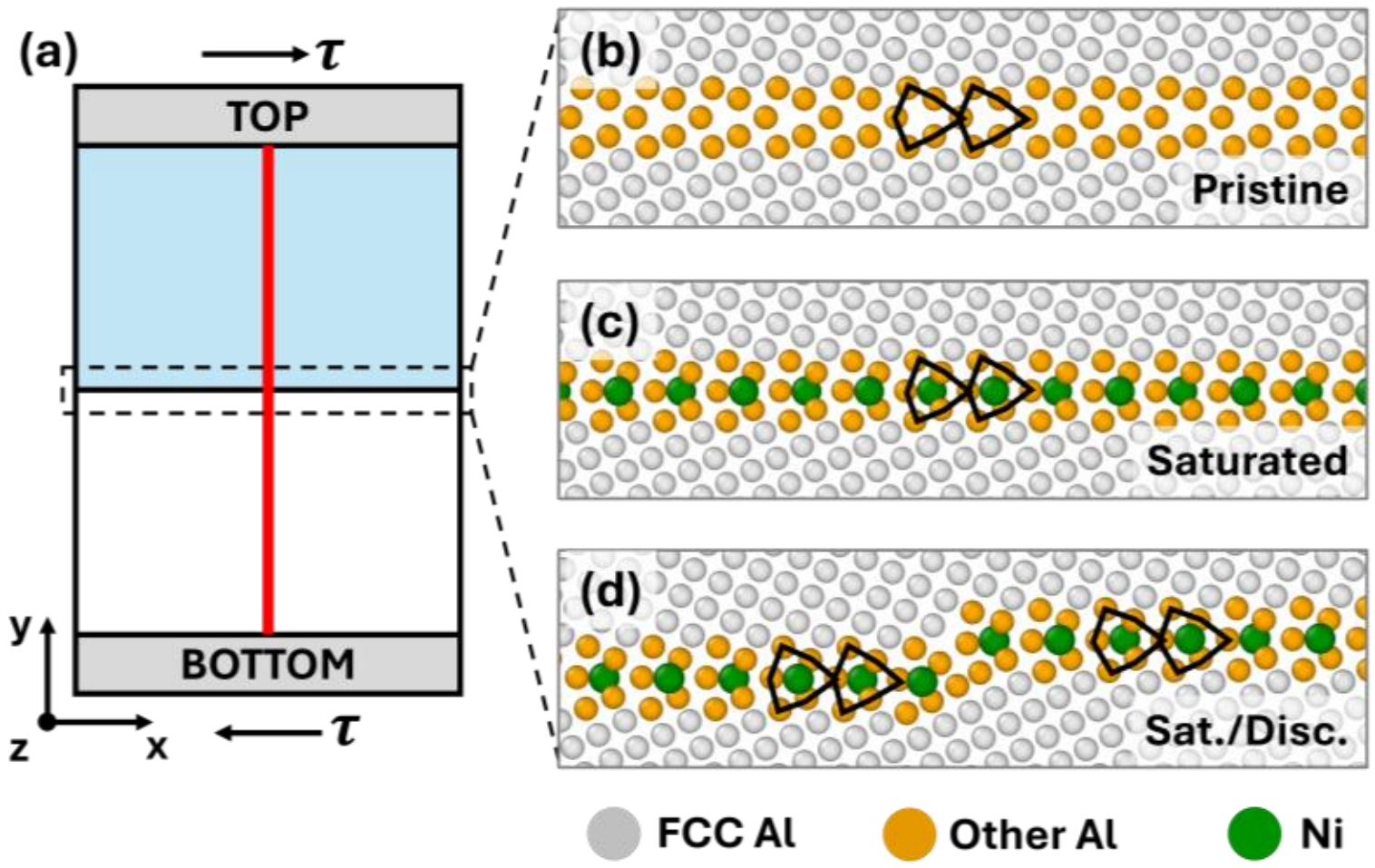

**Fig. S2** (a) Schematic illustration of the shear displacement simulation cell. (b)-(d) Initial GB configurations for the (b) pristine, (c) saturated, and (d) saturated/disconnected (sat./disc.) samples. Ni atoms are enlarged for better visualization.

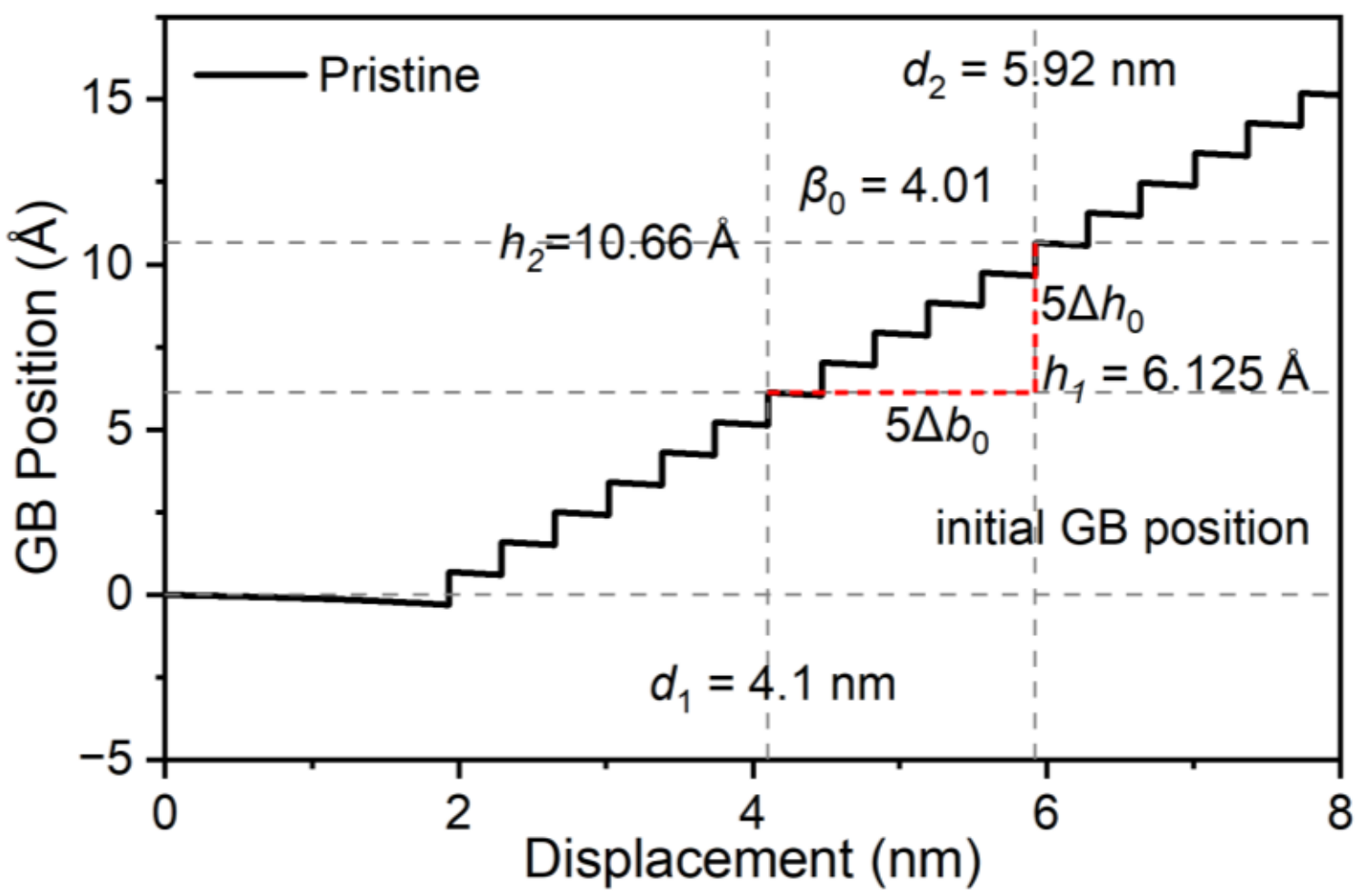

**Fig. S3** GB position vs displacement profile of the pristine sample under applied shear loading. The shear-coupling factor was estimated using $5\beta_0 = |5\Delta b_0/5\Delta h_0|$, which yields $\beta_0 = 4.01$.

### *Estimation of shear-coupling factor*

MS simulations were performed to estimate the magnitude of the shear-coupling factor associated with interstitial segregation-activated GB migration in a Σ5(210) STGB, using the Al-Ni system as a representative case. Following the procedure employed in the first-principles calculations, the 3 and 3' sites were manually substituted with Ni atoms, as shown in Fig. 4(a) and (b), respectively. The bicrystal structure was then relaxed using the conjugate gradient algorithm. During relaxation, the upper and lower layers adjacent to the GB were used as markers. The Burgers vector **b** and step height $h$ were estimated based on the displacement of the mass centers of these two marker regions. Then the shear-coupling factor ($\beta$) was calculated using Eq. (2):

$$\beta = \frac{\Delta x}{\Delta y}. \qquad \text{Eq. (1)}$$

where $\Delta x$ and $\Delta y$ are the relative displacements of the upper and lower marker mass centers along and normal to the GB plane during the interstitial segregation of Ni atoms.

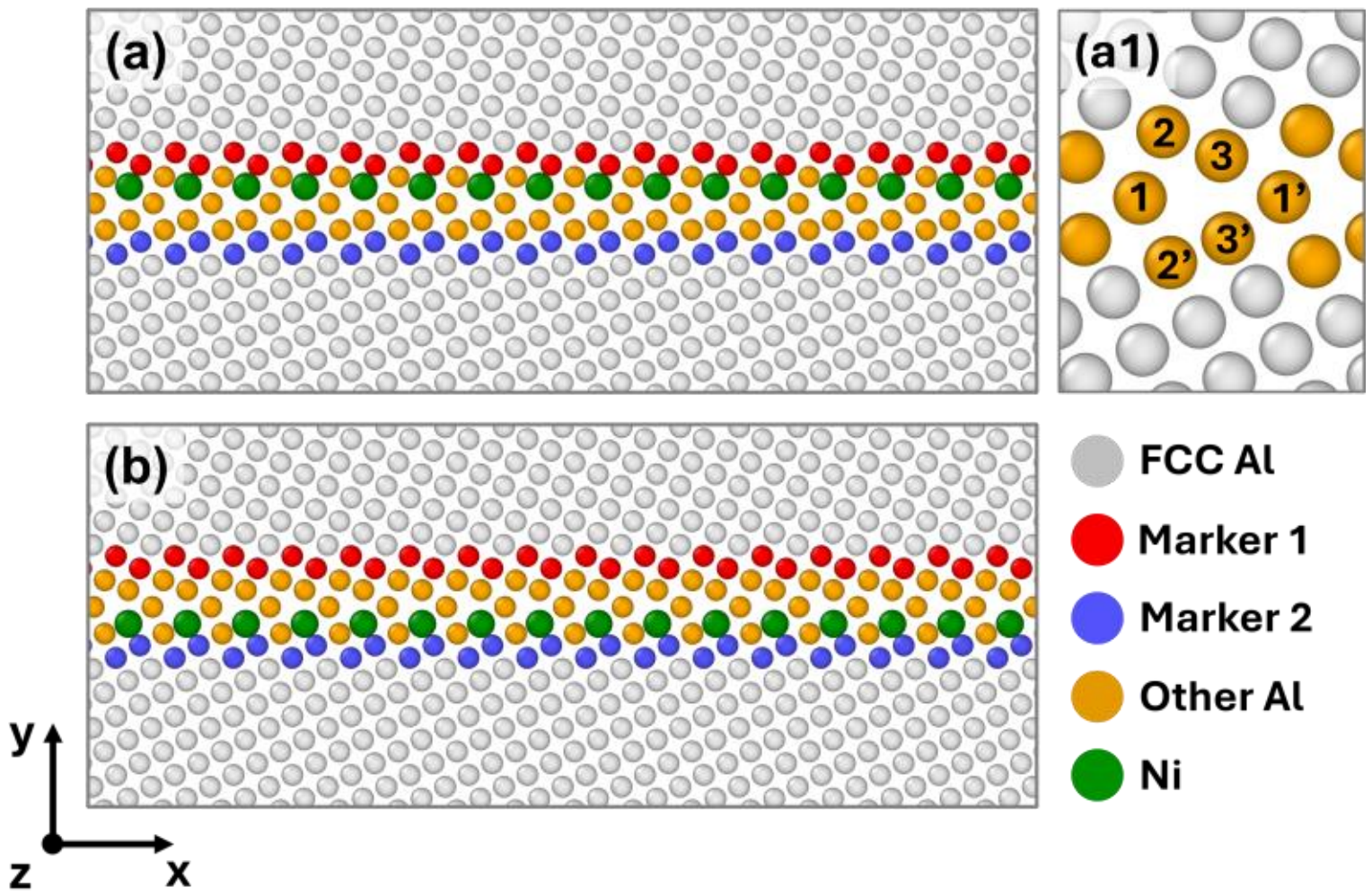

**Fig. S4** Initial Al-Ni Σ5(210) STGB configurations used for shear-coupling factor estimation. In (a), Ni atoms occupy the 3 sites, as indicated in (a1); in (b), Ni atoms substitute the 3′ sites, also shown in (a1). Ni atoms are enlarged for clarity. The red and blue spheres denote Al atoms selected as markers for the shear-coupling factor estimation.

**Table S1** Measured relative displacements of mass centers during the MS simulations. $\beta'$ refers to the estimated shear-coupling factor.

| | $\Delta x$, Å | $\Delta y$, Å | $\beta'$ |
|---|---|---|---|
| Case (a) | 0.95 | 0.75 | 1.27 |
| Case (b) | 0.95 | -0.75 | -1.27 |

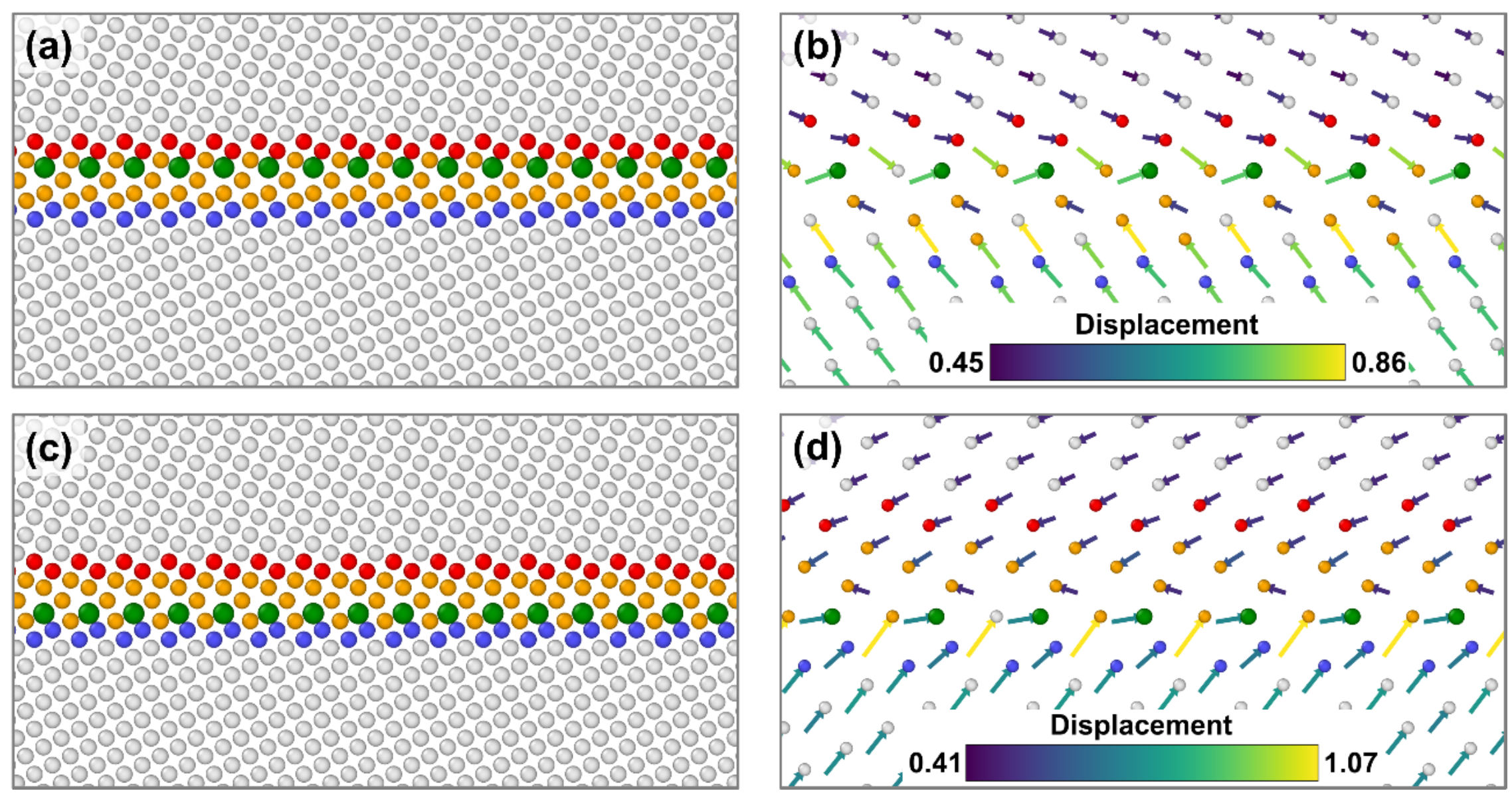

**Fig. S5** (a) and (c) show the initial atomic configurations used for shear-coupling factor estimation for the upward and downward GB migration cases, respectively. (b) and (d) are their corresponding atomic displacement paths and displacement magnitude maps.

Following the MS simulations, the Σ5(210) STGB migrates upward, embracing the Ni atoms at the kite centers, where these solute atoms effectively become interstitials within the kite cores. For this upward-migration case, whose initial structure is shown in Fig. S5(a), in the centers of mass of the upper and lower marker atoms yield a lateral displacement of $\Delta x_1 = 0.95$ Å and a normal migration distance of $\Delta y_1 = 0.75$ Å. Accordingly, the shear-coupling factor is estimated as $\beta_1' = \Delta x_1/\Delta y_1 = 1.27$. An analogous analysis is performed for the downward-migration case, whose initial configuration is shown in Fig. S5(b). The relative displacement of the marker atoms remains $\Delta x_2 = 0.95$ Å, while the migration distance is $\Delta y_1 = -0.75$ Å, where the negative sign indicates downward GB motion. The corresponding shear-coupling factor is therefore $\beta_2' = \Delta x_2/\Delta y_2 = -1.27$. The corresponding atomic displacement paths and displacement magnitude maps are shown in Fig. S4. These results demonstrate that the upward and downward solute interstitial segregation-induced GB migrations exhibit identical magnitudes but opposite signs of the shear-coupling factor.

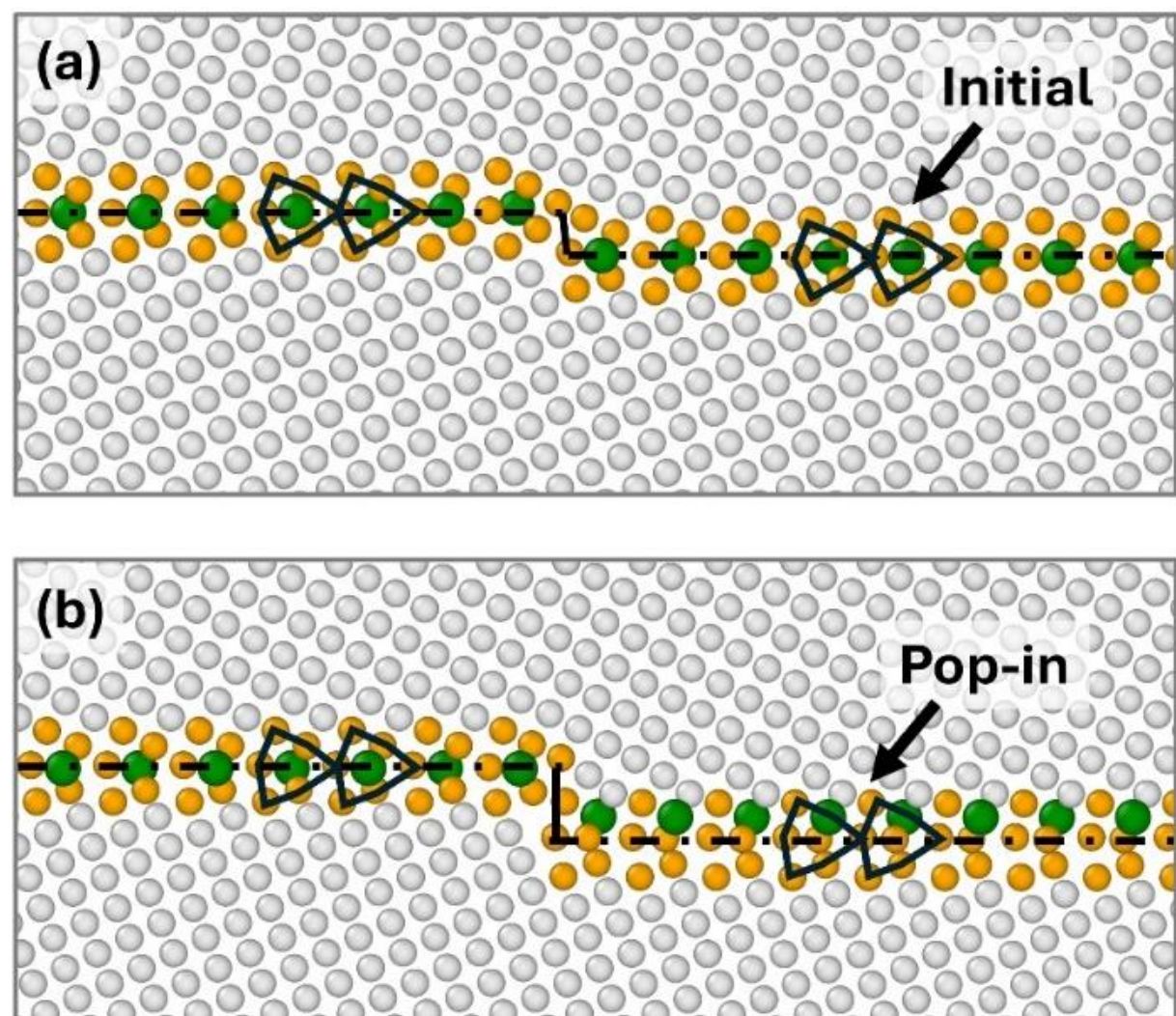

**Fig. S6** (a) Initial GB structure of the Sat./Disc. sample (Al-Ni) when shear displacement d = 0 nm. (b) Deformed GB structure when shear displacement d = 2.5 nm, where part of the kite structures has a one-layer downward migration.

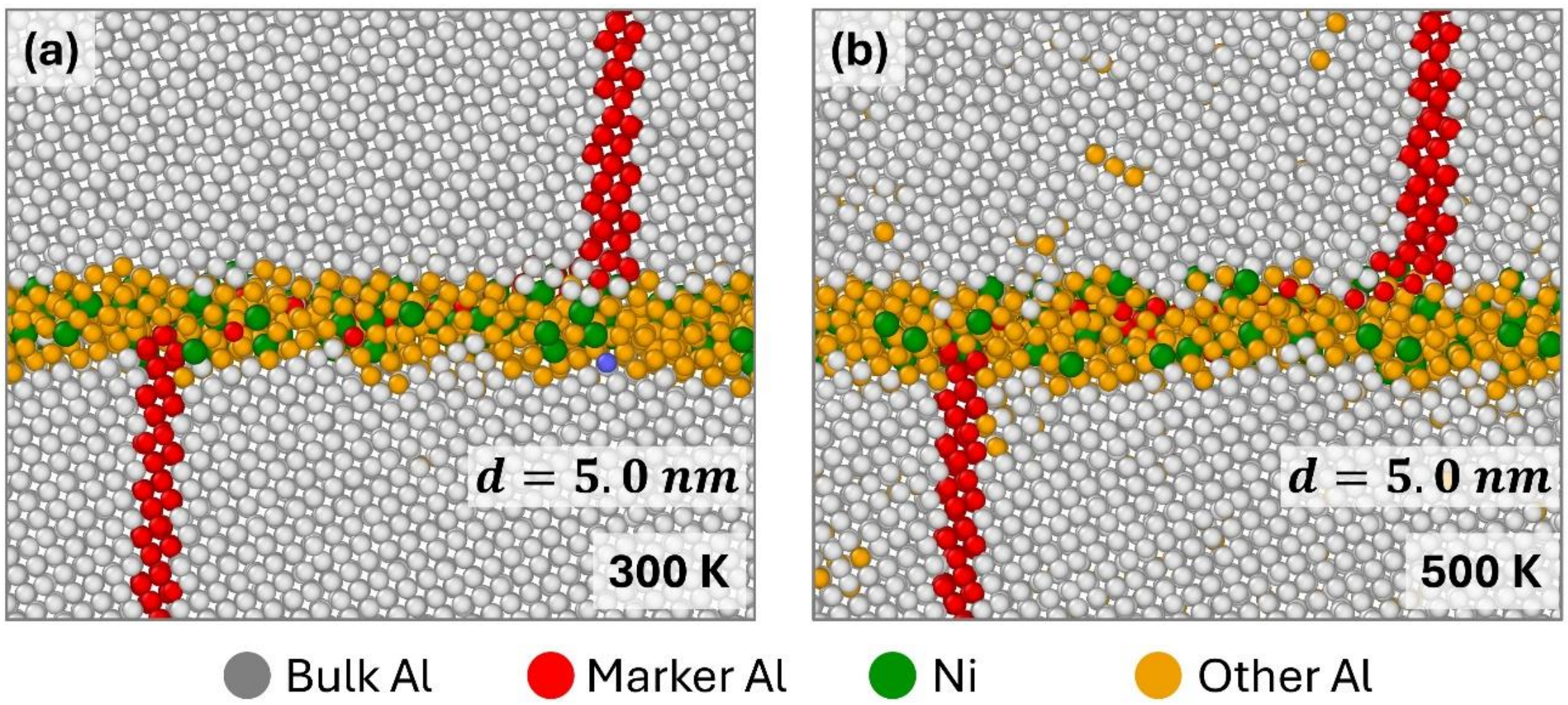

**Fig. S7** Similar to the shear deformation at 0 K, the Sat./Disc. sample also exhibits pure sliding rather than shear-coupled migration under applied shear loading at both 300 K (a) and 500 K (b).

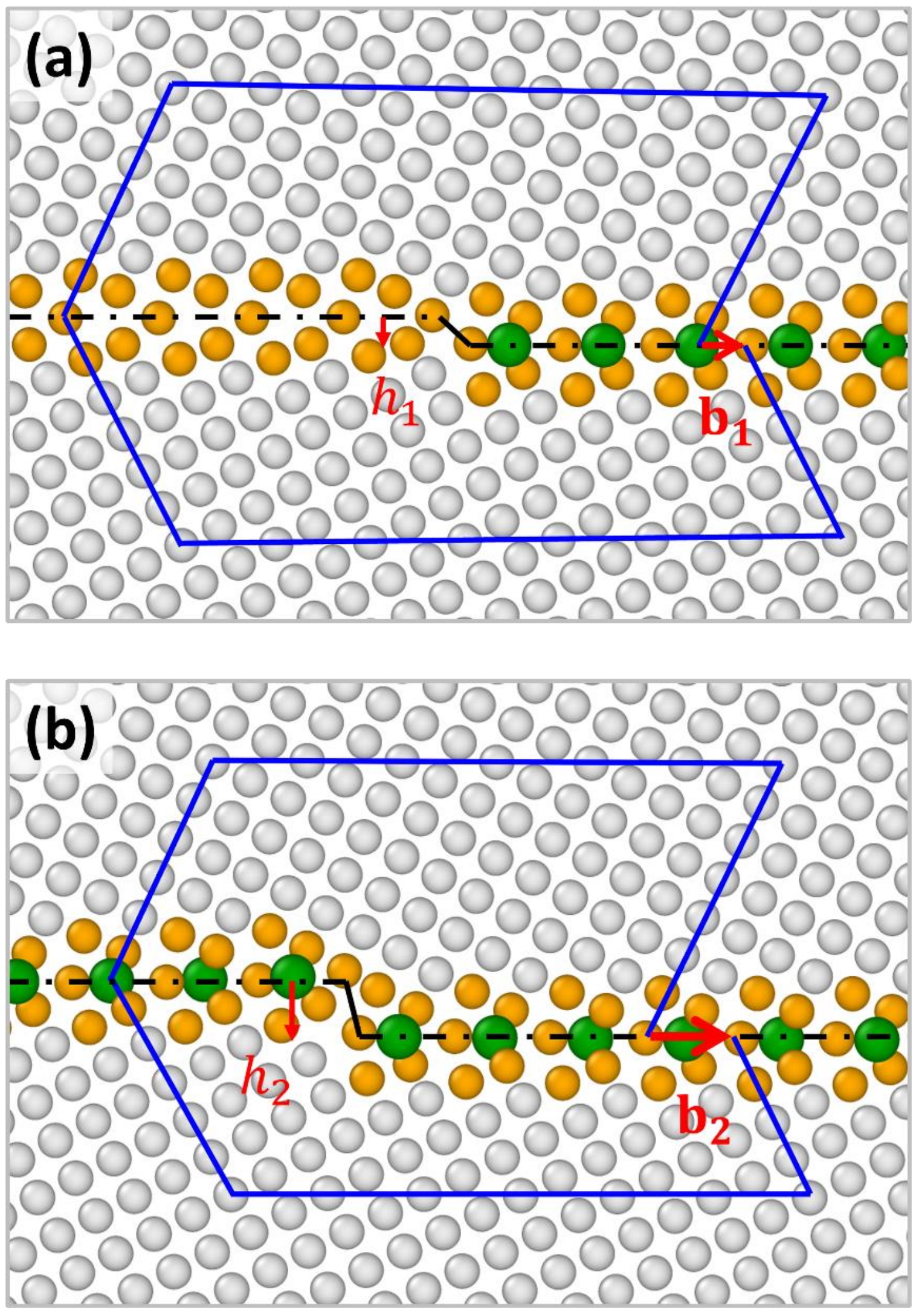

**Fig. S8** Burgers circuit analysis for the (a) the unstable disconnection and (b) the stable disconnection. The step height and Burgers vector of the stable disconnection are twice those of the unstable disconnection.

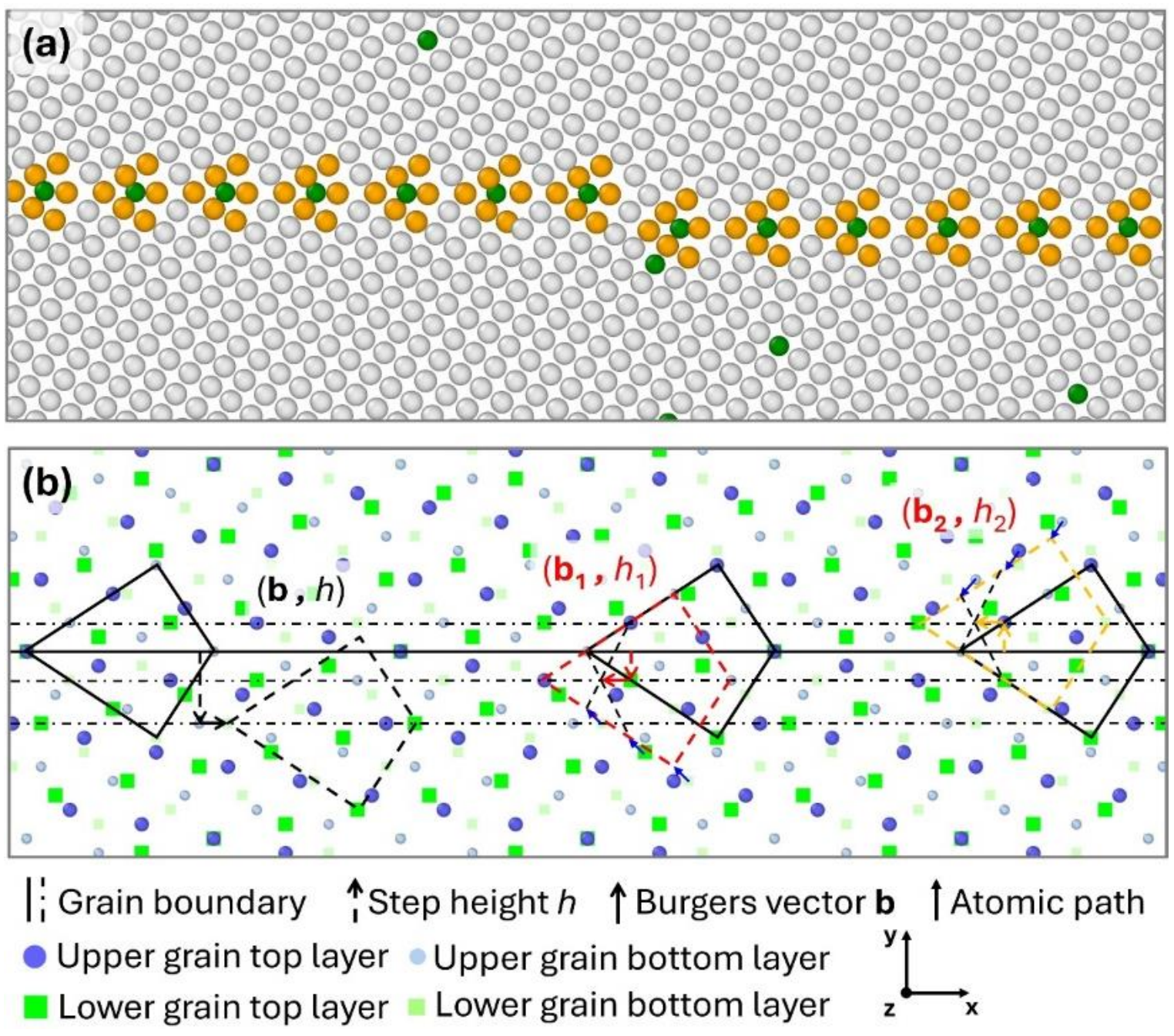

**Fig. S9** (a) Permanent disconnection formed after hybrid MD/MC simulations in the Σ13(320) STGB of the Al-Ni system. (b) Dichromatic pattern analysis of the disconnection nucleation paths in the Σ13(320) STGB. The black dashed kite with (**b**, h) pair indicates the disconnection nucleation path for pristine Σ13(320) STGB, whose shear coupling factor $\beta = b/h = -0.4$ *[3]*. However, the shear coupling factors for segregation activated disconnection are $\beta_1 = -\beta_2 = 1$.

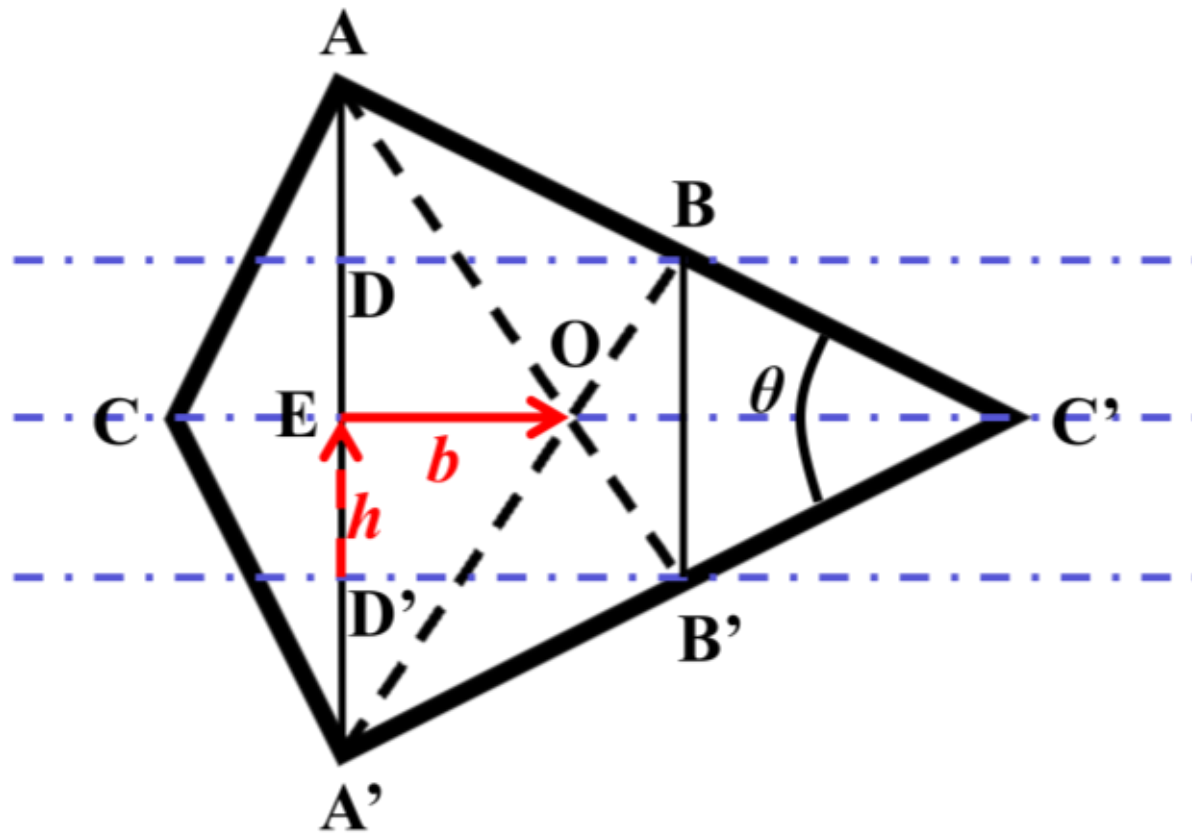

**Fig. S10** Schematic illustration of the nucleation path for a segregation activated disconnection.

***Disconnection nucleation criterion analysis***—In pristine grain boundaries (GBs), classical model predicts that disconnections nucleate in $\langle 100 \rangle$ or $\langle 110 \rangle$ modes depending on the corresponding Burgers vector direction, and resulting in the following nucleation criteria [3]:

$$b_{\langle 100 \rangle} = a \sin(\theta/2), \tag{2}$$

$$h_{\langle 100 \rangle} = (a/2) \cos(\theta/2), \tag{3}$$

$$\beta'_{\langle 100 \rangle} = 2 \tan(\theta/2), \tag{4}$$

and

$$b_{\langle 110 \rangle} = \sqrt{2} a \sin(\varphi/2), \tag{5}$$

$$h_{\langle 110 \rangle} = -\sqrt{2} a \cos(\varphi/2), \tag{6}$$

$$\beta'_{\langle 110 \rangle} = -2 \tan(\varphi/2), \tag{7}$$

where $\theta$ is the tilt angle of the GB, $a$ is the lattice constant for the pure system, and $\varphi$ is the complement angle for $\theta$: $\varphi = 90^{\circ} - \theta$.

Nevertheless, the nucleation pathways for solute segregation–activated disconnections are markedly different from those in pristine tilt GBs, and the corresponding nucleation criteria also differ. To analyze these criteria, we begin with a topological examination of the kite structure in a Σ5(210) STGB. As shown in Fig. S10, points B and B′ are located at the centers of the kite structure and correspond to the midpoints of segments A–C′ and A′–C′, respectively, with each segment having a length equal to the lattice constant $a$. The vertical distance between points A and A′ is always twice the horizontal distance between points B and B', i.e., AA' ≡2BB'. Therefore,

$$b = (a/3)\cos(\theta/2), \tag{8}$$

$$h = (a/2)\sin(\theta/2), \tag{9}$$

$$\beta' = b/h = (2/3)\cot(\theta/2), \tag{10}$$

Eqs. (8) – (10) differ significantly from those in pristine cases. Notably, the direction of the Burgers vectors in segregation-activated disconnections always aligns with the orientation of the kite structure, specifically, along the C–C′ direction, as illustrated in Fig. S6. Therefore, the sign of $\beta$ depends on the migration direction and kite orientation.

## Reference

[1] J.P. Tavenner, M.I. Mendelev, J.W. Lawson, Molecular dynamics based kinetic Monte Carlo simulation for accelerated diffusion, Computational Materials Science 218 (2023) 111929. https://doi.org/10.1016/j.commatsci.2022.111929.

[2] A. Rajabzadeh, F. Mompiou, M. Legros, N. Combe, Elementary Mechanisms of Shear-Coupled Grain Boundary Migration, Phys. Rev. Lett. 110 (2013) 265507. https://doi.org/10.1103/PhysRevLett.110.265507.

[3] J.W. Cahn, Y. Mishin, A. Suzuki, Coupling grain boundary motion to shear deformation, Acta Materialia 54 (2006) 4953–4975. https://doi.org/10.1016/j.actamat.2006.08.004.